\documentclass[reqno,10pt,a4paper,dvips]{amsart}

\usepackage{amssymb,times,mathptmx,cite,eucal,psfrag,array,setspace,geometry,color}
\usepackage[dvips]{graphicx}

\numberwithin{equation}{section}

\newcolumntype{C}{>{$}c<{$}} 

\allowdisplaybreaks

\newcommand{\alg}[1]{\mathfrak{#1}}
\newcommand{\group}[1]{\mathsf{#1}}

\newcommand{\func}[2]{#1 \left( #2 \right)}

\newcommand{\brac}[1]{\left( #1 \right)}
\newcommand{\sqbrac}[1]{\left[ #1 \right]}
\newcommand{\set}[1]{\left\{ #1 \right\}}

\newcommand{\st}{\mspace{5mu} : \mspace{5mu}}

\newcommand{\abs}[1]{\left| #1 \right|}

\newcommand{\ZZ}{\mathbb{Z}}

\newcommand{\dd}{\mathrm{d}}

\newcommand{\diff}[2]{\frac{\dd #1}{\dd #2}}
\newcommand{\pardiff}[2]{\frac{\partial #1}{\partial #2}}

\newcommand{\comm}[2]{\bigl[ #1 , #2 \bigr]}
\newcommand{\dcomm}[2]{\bigl[ \mspace{-4.3mu} \bigl[ #1 , #2 \bigr] \mspace{-4.3mu} \bigr]}

\newcommand{\bra}[1]{\bigl\langle #1 \bigr\rvert}
\newcommand{\ket}[1]{\bigl\lvert #1 \bigr\rangle}
\newcommand{\braket}[2]{\bigl\langle #1 \bigr\rvert \bigl. #2 \bigr\rangle}
\newcommand{\bracket}[3]{\bigl\langle #1 \bigr\rvert #2 \bigl\lvert #3 \bigr\rangle} 
\newcommand{\corrfn}[1]{\bigl\langle #1 \bigr\rangle}

\newcommand{\MinMod}[2]{\mathsf{M} \left( #1 , #2 \right)}
\newcommand{\LogMinMod}[2]{\mathsf{LM} \left( #1 , #2 \right)}
\newcommand{\DualLogMinMod}[2]{\mathsf{L^*M} \left( #1 , #2 \right)}

\newcommand{\VerMod}[1]{\mathcal{V}_{#1}}
\newcommand{\IrrMod}[1]{\mathcal{L}_{#1}}
\newcommand{\IndMod}[1]{\mathcal{M}_{#1}}
\newcommand{\LogMod}[1]{\mathcal{I}_{#1}}
\newcommand{\JorMod}[1]{\mathcal{J}_{#1}}

\newcommand{\fuse}{\times_{\! f}}

\newcommand{\partnum}[1]{\func{P}{#1}}

\newcommand{\eqnref}[1]{Equation~(\ref{#1})}
\newcommand{\eqnDref}[2]{Equations~(\ref{#1}) and (\ref{#2})}
\newcommand{\eqnsref}[2]{Equations~(\ref{#1} -- \ref{#2})}
\newcommand{\secref}[1]{Section~\ref{#1}}
\newcommand{\appref}[1]{Appendix~\ref{#1}}
\newcommand{\figref}[1]{Figure~\ref{#1}}
\newcommand{\tabref}[1]{Table~\ref{#1}}

\newcommand{\cft}{conformal field theory}
\newcommand{\cfts}{conformal field theories}
\newcommand{\ope}{operator product expansion}
\newcommand{\opes}{operator product expansions}
\newcommand{\hws}{highest weight state}
\newcommand{\hwss}{highest weight states}
\newcommand{\hwm}{highest weight module}
\newcommand{\hwms}{highest weight modules}
\newcommand{\hwsm}{highest weight submodule}
\newcommand{\hwsms}{highest weight submodules}
\newcommand{\ode}{ordinary differential equation}

\newcommand{\pde}{partial differential equation}

\DeclareMathOperator{\id}{id}

\newtheorem{proposition}{Proposition}		

\begin{document}

\title[Logarithmic $\MinMod{2}{p}$ Models, Couplings, and Duality]{Logarithmic $\MinMod{2}{p}$ Minimal Models, \\ their Logarithmic Couplings, and Duality}

\author[P Mathieu]{Pierre Mathieu}

\address[Pierre Mathieu]{
D\'{e}partement de Physique, de G\'{e}nie Physique et d'Optique \\
Universit\'{e} Laval \\
Qu\'{e}bec, Canada G1K 7P4
}

\email{pmathieu@phy.ulaval.ca}

\author[D Ridout]{David Ridout}

\address[David Ridout]{
Theory Group, DESY \\
Notkestra\ss{}e 85 \\
D-22603, Hamburg, Germany
}

\email{dridout@mail.desy.de}

\thanks{\today \\ This work is supported by NSERC and the Marie Curie Excellence Grant MEXT-CT-2006-042695.}

\begin{abstract}
A natural construction of the logarithmic extension of the $\MinMod{2}{p}$ (chiral) minimal models is presented, which generalises our previous model \cite{RidPer07} of percolation ($p=3$).  Its key aspect is the replacement of the minimal model irreducible modules by reducible ones obtained by requiring that only one of the two principal singular vectors of each module vanish.  The resulting theory is then constructed systematically by repeatedly fusing these building block representations.  This generates indecomposable representations of the type which signify the presence of logarithmic partner fields in the theory.  The basic data characterising these indecomposable modules, the logarithmic couplings, are computed for many special cases and given a new structural interpretation.  Quite remarkably, a number of them are presented in closed analytic form (for general $p$).  These are the prime examples of ``gauge-invariant'' data --- quantities independent of the ambiguities present in defining the logarithmic partner fields.  Finally, mere global conformal invariance is shown to enforce strong constraints on the allowed spectrum:  It is not possible to include modules other than those generated by the fusion of the model's building blocks.  This generalises the statement that there cannot exist two effective central charges in a $c=0$ model.  It also suggests the existence of a second ``dual'' logarithmic theory for each $p$.  Such dual models are briefly discussed.
\end{abstract}

\maketitle

\onehalfspacing

\section{Introduction} \label{secIntro}

Numerical simulations of non-local observables in simple two-dimensional critical statistical systems --- for example, crossing probabilities for percolation or the Ising model --- have revealed a surprising result:  These non-local observables require the presence of non-unitary representations lying outside the Kac table of the corresponding minimal model \cft{} \cite{LanCon94,LapCro01,ArgNon02}.

To make the discussion of this result transparent, let us recall that the minimal models $\MinMod{p'}{p}$ are parametrised by two coprime positive integers $p$ and $p'$ (with say $p>p'$), and have central charge
\begin{equation}
c_{p',p} = 1 - \frac{6 \brac{p-p'}^2}{p p'}.
\end{equation}
The (chiral) primary fields which populate these models will be denoted by $\phi_{r,s}$, for $r = 1 , 2 , \ldots , p'-1$ and $s = 1 , 2 , \ldots ,
p-1$, and have conformal dimensions
\begin{equation} \label{eqnConfDim}
h_{r,s} = \frac{\brac{pr - p's}^2 - \brac{p-p'}^2}{4 p p'}.
\end{equation}
These dimensions are conveniently arranged into the \emph{Kac table} of the minimal model.

Critical percolation is described by a $c=0$ theory, hence would correspond, by
a na\"{\i}ve central charge identification, to the $\MinMod{2}{3}$ model.  However, the horizontal crossing probability for this theory (which can be roughly identified with a four-point function of $\phi_{1,2}$ \cite{CarCri92}) indirectly indicates the presence of a field of dimension $h_{1,3} = \tfrac{1}{3}$, which lies outside the $\MinMod{2}{3}$ Kac table.  (Of course, the necessity of going beyond, in some way, the $\MinMod{2}{3}$ theory is actually a plain consequence of the triviality of this model.)  Similarly, crossing probabilities for the Ising model involve a field of dimension $h_{3,3} = \frac{1}{6}$, which again lies outside the $\MinMod{3}{4}$ Kac table \cite{LapCro01}.  Additional non-local observables have been probed for the Ising model in \cite{ArgNon02}, where in particular, a field of dimension $h_{3,1} = \tfrac{5}{3}$ appears to be relevant.

These results suggest the existence of some sort of covering theory for a general minimal model which, contrary to its minimal model reduction, is not blind to non-local observables.  The systematic construction of such a generalisation, for the specific minimal model related to percolation theory, was considered in \cite{RidPer07}.  The resulting theory turned out to be a chiral logarithmic \cft{}, which we will refer to as the \emph{logarithmic chiral minimal model} $\LogMinMod{2}{3}$ to stress that it generalises (in a manner we consider very natural) the chiral part of the (trivial) minimal model $\MinMod{2}{3}$.  This logarithmic signature was marked by the presence of indecomposable representations that are generated by fusing the building block representations of critical percolation.  We will review our construction briefly in \secref{secLM23}.

As a brief aside, we would like to emphasise that many of the non-local observables that are typically considered in this context, crossing probabilities in particular, are necessarily defined in the presence of \emph{boundaries}.  This then places the theoretical formalism for describing these observables within the realm of \emph{boundary} \cft{} \cite{CarEff86}.  In other words, any theory describing such non-local observables is intrinsically chiral.  This explains why the logarithmic theory that we have constructed in \cite{RidPer07} is chiral, as are the theories we propose here.  Of course, we think that it is desirable that these chiral logarithmic \cfts{} admit ``lifts'' to consistent (modular invariant) bulk theories.  However, we do not view this requirement as mandatory, and it could very well be that such a lift does not exist.  Our point is that within the context considered here, it is the boundary \cft{} which has direct physical meaning, and therefore this chiral theory must itself be consistently defined\footnote{We emphasise that we have not proven that the description we give of this chiral theory is complete.  Rather, we content ourselves by noting that we have closure under fusion and consistency with global conformal invariance (pointing out that the latter rules out the most obvious proposed extensions of our description).  We expect to return to this issue of the characterisation of a consistent chiral theory in future work.}.

The proposal that every minimal model might be ``augmented'' in some way to define a corresponding logarithmic model has been suggested recently by various authors \cite{FjeLog02,EbeVir06,FeiLog06,PeaLog06,ReaAss07,RasFus07}.  Our proposal differs from all of these, although it inevitably shares a number of features with them (this is most visible with those which are also intrinsic constructions at the level of the Virasoro algebra).  In particular, the motivation underlying the construction of the lattice models (formulated in terms of non-local variables) in \cite{PeaLog06} is quite similar to ours.  For the specific logarithmic extension of the $\MinMod{2}{3}$ model, the differences between our construction and those of \cite{EbeVir06,ReaAss07,RasFus07} are spelled out in the conclusion of \cite{RidPer07}.

The aim of the present work is to generalise this construction by presenting a logarithmic chiral version of all the $\MinMod{2}{p}$ models for $p$ odd.  In \secref{secLM23}, we first briefly review the results of \cite{RidPer07}, with a
special emphasis on the fundamental data characterising certain indecomposable representations (staggered modules \cite{RohRed96}) that appear, namely the logarithmic couplings.  These quantify the linking of the constituent modules that comprise these representations.  A further characterisation of these couplings is presented here, in terms of singular vectors.  This new observation explains why these couplings have particular values (they are \emph{not} free parameters), and provides a completely transparent method for computing them.  We also stress the notion of ``gauge transformations'', reflecting the degrees of freedom inherent in the definition of logarithmic partner states, and hence the importance of gauge-invariant data.

The $\LogMinMod{2}{5}$ model is then considered in some detail in \secref{secLM25}.  We give several fusion rules explicitly, as well as the values of the first few logarithmic couplings.  In \secref{secLM2p}, these results are extended to all $\MinMod{2}{p}$ models with $p$ odd.  Quite amazingly, we are able to compute a number of logarithmic couplings for general $p$, obtaining remarkably simple closed analytic forms.  This leads to a conjecture for the general form of certain logarithmic couplings, in particular for the modules which extend the vacuum module.  Non-trivial computational evidence is given to support this conjecture.

Next, a constraint induced by nothing more than global conformal invariance is shown, in \secref{secInconsistency}, to prevent our na\"{\i}ve attempts to extend the spectra of our $\LogMinMod{2}{p}$ models beyond that which we have considered.  This generalises a result already presented in our previous analysis of percolation \cite{RidPer07} ($p=3$) for which the argument boils down to the statement that the effective central charge (which is a distinguished logarithmic coupling constant) of a $c=0$ theory is unique \cite{GurCon02,GurCon04}.

This constraint then suggests the existence of a \emph{dual} logarithmic theory $\DualLogMinMod{2}{p}$ (for each $p$ odd) which shares the same central charge as $\LogMinMod{2}{p}$, but which has a completely different spectrum (only the vacuum module is common to both).  We briefly discuss these dual models in \secref{secDual}, exhibiting their fusion rules and some of their logarithmic couplings.  

We finish with a brief summary of our results and some conclusions.  This is followed by two technical appendices which justify (and clarify) certain computations used to derive the spectrum-limiting constraint of \secref{secInconsistency}.  These appendices may be of independent interest, and should also serve to allay any suspicion that this constraint might be circumvented in some way.  They are followed by a third appendix which clarifies the physical significance of the choice of inner product on the indecomposable modules which appear in these logarithmic theories.  We stress that this differs from the conventional choice sometimes advocated in the logarithmic \cft{} literature\footnote{The literature on this subject is now rather vast.  There are at least three main sources of logarithmic models:  Wess-Zumino-Witten models with supergroup symmetries \cite{RozQua92,SalGL106,GotWZN07}, models with an affine Lie symmetry algebra at fractional level \cite{GabFus01,LesLog04}, and non-minimal models of the type $\MinMod{1}{p}$ \cite{GurLog93,GabLoc99}.  Standard reviews which emphasise the latter two classes of models are \cite{GabAlg03,FloBit03}.  The scalar product used by Flohr and collaborators \cite{FloSin98,FloBit03,FloNot06} differs from our choice.  However, our convention does agree with that of Gurarie and Ludwig \cite{GurCon02,GurCon04}.}.

\section{Notation} \label{secNotation}

As we have already noted, it is customary to represent the dimensions of the primary fields of a minimal model $\MinMod{p'}{p}$ in a table, the Kac table of the theory.  The \emph{Kac symmetry} $h_{r,s} = h_{p'-r,p-s}$ indicates field identifications within this theory.  We will be more interested in the table of dimensions obtained from \eqnref{eqnConfDim} by relaxing the conditions on $r$ and $s$ to $r,s \in \ZZ_+$.  We refer to this table as the \emph{extended Kac table}, and use the same notation $\phi_{r,s}$ to denote primary fields whose dimension is given by \eqnref{eqnConfDim} (for arbitrary $r,s \in \ZZ_+$).

We will denote the Verma module generated from the \hws{} $\ket{\phi_{r,s}}$ by $\VerMod{r,s}$ and its irreducible quotient by $\IrrMod{r,s}$.  Note that at central charge $c = c_{p',p}$, the Verma module $\VerMod{r,s}$ with $r$ divisible by $p'$ or $s$ divisible by $p$ (in brief, $p' \mid r$ or $p \mid s$) has a maximal submodule generated by a single singular vector.  In contrast, the maximal submodules of the other $\VerMod{r,s}$ associated to the extended Kac table are generated by two singular vectors \cite{FeiVer84}.  The modules $\IndMod{r,s}$ which form the primary focus of our investigations are obtained however by quotienting each $\VerMod{r,s}$ by the Verma module generated by the singular vector at grade $rs$.  Under some circumstances, specifically when $r = p'$ and $s \leqslant p$ or $s = p$ and $r \leqslant p'$, this yields the irreducible module:  $\IndMod{r,s} = \IrrMod{r,s}$.  The other $\IndMod{r,s}$ are however reducible but indecomposable\footnote{We recall that a module is \emph{reducible} if it contains a non-trivial submodule and \emph{decomposable} if it can be written as the direct sum of two non-trivial submodules.}.

\section{$\LogMinMod{2}{3}$:  Critical Percolation} \label{secLM23}

Our construction of the $\LogMinMod{2}{3}$ logarithmic theory is most simply viewed as a modification of the (chiral) $\MinMod{2}{3}$ minimal model.  The latter model is composed of two irreducible modules $\IrrMod{1,1}$ and $\IrrMod{1,2}$ which are identified by the Kac symmetry.  The modification to $\MinMod{2}{3}$ consists of breaking this symmetry in a specific way, by replacing the two irreducible modules with their reducible (and non-isomorphic) counterparts $\IndMod{1,1}$ and $\IndMod{1,2}$.  This means that in each module, one of the two principal singular vectors is not set equal to zero, although it still has zero norm.  Specifically, the singular vector at level $2$ in $\VerMod{1,1}$ (which corresponds to the energy-momentum tensor) and the one at level $1$ in $\VerMod{1,2}$ are no longer vanishing (they are said to be physical).  This is how the Kac symmetry is broken and we stress that this is our sole input to the formulation of $\LogMinMod{2}{3}$.

In the context of percolation theory, this particular construction is supported by firm physical considerations:  This is the minimal way in which we can modify $\MinMod{2}{3}$ so as to generate a theory that is consistent with Cardy's computation \cite{CarCri92} of the horizontal crossing probability for critical percolation.  The module $\IndMod{1,2}$, the central object in Cardy's theory, is the building block of our model.  In particular, the \emph{spectrum}, that is, the set of modules from which the model is composed, appears to be completely determined by repeatedly fusing the module $\IndMod{1,2}$ with itself (see \secref{secInconsistency}).

\begin{table}
\begin{center}
\setlength{\extrarowheight}{4pt}
\begin{tabular}{|C|C|C|C|C|C|C|C|C|C|C}
\hline
0 & 0 & \tfrac{1}{3} & 1 & 2 & \tfrac{10}{3} & 5 & 7 & \tfrac{28}{3} & 12 & \cdots \\[1mm]
\hline
\tfrac{5}{8} & \tfrac{1}{8} & \tfrac{-1}{24} & \tfrac{1}{8} & \tfrac{5}{8} & \tfrac{35}{24} & \tfrac{21}{8} & \tfrac{33}{8} & \tfrac{143}{24} & \tfrac{65}{8} & \cdots \\[1mm]
\hline
2 & 1 & \tfrac{1}{3} & 0 & 0 & \tfrac{1}{3} & 1 & 2 & \tfrac{10}{3} & 5 & \cdots \\[1mm]
\hline
\end{tabular}
\vspace{3mm}
\caption{The first three rows of the extended Kac table for $c=c_{2,3}=0$, listing the dimensions $h_{r,s}$ of the primary fields $\phi_{r,s}$.  Here, $r$ increases downwards, and $s$ increases to the right, so the top-left-hand corner corresponds to the identity field $\phi_{1,1}$, which with $\phi_{1,2}$ exhausts the Kac table of $\MinMod{2}{3}$.} \label{tabExtKacc=0}
\end{center}
\end{table}

In \cite{RidPer07}, we computed the fusion rules of our theory using the algorithm\footnote{We implemented this algorithm in \textsc{Maple 10}.} of Nahm and Gaberdiel-Kausch \cite{NahQua94,GabInd96}, which is completely algebraic (making no reference to correlators and differential equations) and distinguishes between vanishing and non-vanishing singular vectors.  The modules that are generated by these fusions can all be described in terms of the top row  of the extended Kac table for $c=0$.  We display a part of this table in \tabref{tabExtKacc=0} (restricted here to $r=1,2,3$ and $s=1,\ldots,10$).  

These fusion rules imply that the spectrum of $\LogMinMod{2}{3}$ must consist of at least
\begin{equation} \label{eqnSpec23}
\set{\IndMod{1,s} \st 3 \mid s \geqslant 3} \cup \set{\LogMod{1,s} \st 3 \nmid s > 3}.
\end{equation}
Here the modules $\LogMod{1,s}$ denote staggered modules \cite{RohRed96} of rank $2$.  As a vector space, $\LogMod{1,s}$ is isomorphic to $\IndMod{1,s'} \oplus \IndMod{1,s}$, where
\begin{equation} \label{eqnHWSMLabel23}
s' = 
\begin{cases}
s-2 & \text{if $s = 1 \pmod{3}$,} \\
s-4 & \text{if $s = 2 \pmod{3}$,}
\end{cases}
\end{equation}
but this is \emph{not} an isomorphism of $\alg{Vir}$-modules\footnote{However, the vector space isomorphism implies that the characters of $\LogMod{1,s}$ and $\IndMod{1,s'} \oplus \IndMod{1,s}$ are identical.}.  The $\LogMod{1,s}$ are in fact reducible, but indecomposable, modules with a maximal \hwsm{} isomorphic to $\IndMod{1,s'}$.  Furthermore, quotienting $\LogMod{1,s}$ by the submodule $\IndMod{1,s'}$ gives the \hwm{} $\IndMod{1,s}$.  We mention that every $\IndMod{1,s}$ with $3 \nmid s$ appears as a \hwsm{} of one of these staggered modules, so we can restrict our attention to these submodules when appropriate.  For example, the vacuum module $\IndMod{1,1}$ appears in this way as the \hwsm{} of $\LogMod{1,5}$.
  
Although rather difficult to calculate (for all but the most trivial cases, a computer is required), the fusion rules of $\LogMinMod{2}{3}$ can be expressed in a rather elegant and natural way.  This description of the fusion rules makes use of the ``auxiliary rule''
\begin{equation} \label{eqnMFuseM}
\IndMod{1,s} \fuse \IndMod{1,t} = \IndMod{1,\abs{s-t}+1} \oplus \IndMod{1,\abs{s-t}+3} \oplus \ldots \oplus \IndMod{1,s+t-3} \oplus \IndMod{1,s+t-1},
\end{equation}
which we stress does not itself give correct results.  Instead, the correct fusion rules (including those involving staggered modules $\LogMod{1,s}$) are computed using the following simple procedure:
\begin{enumerate}
\item Replace any $\LogMod{1,s}$ by the direct sum $\IndMod{1,s'} \oplus \IndMod{1,s}$ (with $s'$ given by \eqnref{eqnHWSMLabel23}).
\item Compute the ``fusion'' using distributivity and \eqnref{eqnMFuseM}.
\item In the result, replace all direct sums of the form $\IndMod{1,s'} \oplus \IndMod{1,s}$ by $\LogMod{1,s}$ (there is only ever one way to consistently do this\footnote{Actually this uniqueness condition only holds when we fuse fully extended modules --- those appearing in (\ref{eqnSpec23}).  For example, in fusing $\IndMod{1,5}$ with itself, it is not clear whether $\IndMod{1,5}$ couples to $\IndMod{1,1}$ or $\IndMod{1,7}$.}).
\end{enumerate}
In other words, we compute the fusion of indecomposable modules by fusing at the level of vector spaces, and then reconstructing the module structure via the above uniqueness condition.

\begin{figure}
\begin{center}
\includegraphics[width=5cm]{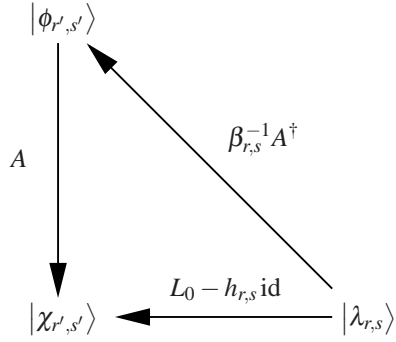}
\caption{The relationship between the generating states of a general rank $2$ staggered module $\LogMod{r,s}$.  Here, $\ket{\phi_{r',s'}}$ is the \hws{} from which the non-vanishing singular vector $\ket{\chi_{r',s'}}$ is obtained by acting with the composite Virasoro operator $A$, $\ket{\lambda_{r,s}}$ is the Jordan partner of $\ket{\chi_{r',s'}}$ with dimension $h_{r,s}$, and $\beta_{r,s}$ is the logarithmic coupling constant of $\LogMod{r,s}$.} \label{figStagMod}
\end{center}
\end{figure}

The logarithmic nature of the theory is due to the non-diagonalisability of $L_0$ on the staggered modules $\LogMod{1,s}$.  Every state in $\LogMod{1,s}$ can be realised as a descendant of one of \emph{two} generating states $\ket{\phi_{1,s'}}$ and $\ket{\lambda_{1,s}}$.  The generator $\ket{\phi_{1,s'}}$ is a \hws{} of dimension $h_{1,s'}$ (with $s'$ given by \eqnref{eqnHWSMLabel23}) which generates a module isomorphic to the indecomposable submodule $\IndMod{1,s'}$.  There is a non-vanishing singular vector $\ket{\chi_{1,s'}}$ descended from $\ket{\phi_{1,s'}}$, and its dimension is $h_{1,s}$.  The other generating state $\ket{\lambda_{1,s}}$ is now realised as the \emph{Jordan partner} to $\ket{\chi_{1,s'}}$ in a rank $2$ Jordan cell, and may be normalised such that
\begin{equation} \label{eqnDefLambda}
L_0 \ket{\lambda_{1,s}} = h_{1,s} \ket{\lambda_{1,s}} + \ket{\chi_{1,s'}}.
\end{equation}
Here we must also choose a normalisation for $\ket{\chi_{1,s'}}$.  We illustrate this structure (quite generally) in \figref{figStagMod}.

\psfrag{phi}[][]{$\ket{\phi_{r',s'}}$}
\psfrag{chi}[][]{$\ket{\chi_{r',s'}}$}
\psfrag{lam}[][]{$\ket{\lambda_{r,s}}$}
\psfrag{A}[][]{$A$}
\psfrag{B}[][]{$\beta_{r,s}^{-1} A^{\dag}$}
\psfrag{L}[][]{$L_0 - h_{r,s} \id$}

The generator $\ket{\lambda_{1,s}} \in \LogMod{1,s}$ is not primary, although its image in the quotient space $\IndMod{1,s} = \LogMod{1,s} / \IndMod{1,s'}$ is.  We therefore must have $L_n \ket{\lambda_{1,s}} \in \IndMod{1,s'}$ for all $0 < n \leqslant h_{1,s} - h_{1,s'}$.  Determining this action of the positive Virasoro modes on $\ket{\lambda_{1,s}}$ is the fundamental prerequisite for being able to compute in $\LogMod{1,s}$.  This contrasts with the familiar case of a \hwm{}, in which the action of the positive modes can be deduced from that of the non-positive modes and the commutation relations.

The Nahm-Gaberdiel-Kausch fusion algorithm may be used to calculate this positive mode action in specific cases, but, as we remarked in \cite{RidPer07}, this action is not well-defined in general.  The issue is that the Jordan partner condition (\eqnref{eqnDefLambda}) does not determine $\ket{\lambda_{1,s}}$ completely:  It is invariant under \emph{gauge transformations}\footnote{We refer to these as gauge transformations even though they do not depend on a point in space.  One could argue however that they are localised in the state space in that they differ from one module to another.  The basic concepts of gauge theory apply to quite general quotient constructions, and we feel that they provide a convenient and familiar language with which to understand the subtleties of staggered modules.} of the form
\begin{equation} \label{eqnGTlambda}
\ket{\lambda_{1,s}} \longrightarrow \ket{\lambda_{1,s}} + \ket{\psi},
\end{equation}
where $\ket{\psi}$ is any $L_0$-eigenstate of dimension $h_{1,s}$ (that is, $\ket{\psi} \in \ker \brac{L_0 - h_{1,s} \id}$).  If there is only one such eigenstate (up to scalar multiples), then it is $\ket{\chi_{1,s'}}$, and the action of the positive Virasoro modes on $\ket{\lambda_{1,s}}$ is then well-defined (\emph{gauge-invariant}) as $\ket{\chi_{1,s'}}$ is a \hws{}.  In general however, $\LogMod{1,s}$ will contain $L_0$-eigenstates of dimension $h_{1,s}$ which are not \hwss{}, so $L_n \ket{\lambda_{1,s}}$ (for $n>0$) will not be gauge-invariant.

Nevertheless, there is a combination of positive Virasoro modes whose action on $\ket{\lambda_{1,s}}$ is gauge-invariant.  Let $A$ be the operator composed of negative Virasoro modes for which $A \ket{\phi_{1,s'}} = \ket{\chi_{1,s'}}$.  Then, $A^{\dag}$ is composed of positive Virasoro modes and $A^{\dag} \ket{\lambda_{1,s}} = \beta_{1,s} \ket{\phi_{1,s'}}$ is gauge-invariant, because
\begin{equation} \label{eqnDefInnProd}
\bracket{\phi_{1,s'}}{A^{\dag}}{\psi} = \braket{\chi_{1,s'}}{\psi} = 0 \qquad \text{for all $\ket{\psi} \in \ker \brac{L_0 - h_{1,s} \id} \subset \IndMod{1,s'}$.}
\end{equation}
The constant
\begin{equation} \label{eqnDefLogCoupling}
\beta_{1,s} = \braket{\chi_{1,s'}}{\lambda_{1,s}}
\end{equation}
is the essential characteristic of the staggered module $\LogMod{1,s}$, and depends only\footnote{It is not hard to see that if we scale $\ket{\chi_{1,s'}}$ by a factor of $a$, then $\beta_{1,s}$ scales by a factor of $\abs{a}^2$.} upon the chosen normalisation of the singular vector $\ket{\chi_{1,s'}}$.  Unless indicated to the contrary, we will always assume the normalisation
\begin{equation} \label{eqnSVNorm}
\ket{\chi_{1,s'}} = \brac{L_{h_{1,s'} - h_{1,s}} + \ldots} \ket{\phi_{1,s'}},
\end{equation}
where the omitted terms are each ordered so that the mode indices in each term are increasing (Poincar\'{e}-Birkhoff-Witt order).  We call the gauge-invariant $\beta_{1,s}$ the \emph{logarithmic coupling} of $\LogMod{1,s}$.

As an aside, we remark that \eqnDref{eqnDefInnProd}{eqnDefLogCoupling} assume that we have defined some sort of inner product on our staggered modules.  We \emph{always} define this inner product so that the \hws{} $\ket{\phi_{1,s'}}$ of the staggered module has norm $1$.  In particular, the vacuum has norm $1$.  Since this differs from some conventions found in the literature (see \cite{FloNot06} for example), we refer the reader to \appref{appInnProd} for a full justification of this choice. 

As the logarithmic coupling may be computed by acting with a certain combination of positive Virasoro modes on $\ket{\lambda_{1,s}}$, the Nahm-Gaberdiel-Kausch algorithm can be used to compute it.  In this way, we found \cite{RidPer07} that
\begin{equation} \label{eqnOldBetas23}
\beta_{1,4} = \frac{-1}{2}, \qquad \beta_{1,5} = \frac{-5}{8} \qquad \text{and} \qquad \beta_{1,7} = \frac{-35}{3}.
\end{equation}
But using this algorithm is not entirely satisfactory as it gives no understanding of why we observe these particular logarithmic couplings, no matter which fusion rules are employed to generate the staggered modules.  Actually, it would seem reasonable, \emph{a priori}, to suppose that one can define a $1$-parameter family of staggered modules which are structurally identical except for the value of this coupling \cite{RohRed96}.

This supposition is however false.  In the quotient module $\IndMod{1,s} = \LogMod{1,s} / \IndMod{1,s'}$, the vanishing states (relative to the Verma module $\VerMod{1,s}$) are generated by a single vanishing singular vector.  This must lift to a vanishing vector in $\LogMod{1,s}$, and the existence of such a lift \emph{uniquely determines} the logarithmic coupling $\beta_{1,s}$.

Let us illustrate this with an example.  In $\LogMod{1,4}$, we have a \hws{} $\ket{\phi_{1,2}}$ of dimension $0$ with non-vanishing singular descendant $\ket{\chi_{1,2}} = L_{-1} \ket{\phi_{1,2}}$.  Its Jordan partner therefore satisfies
\begin{equation}
L_0 \ket{\lambda_{1,4}} = \ket{\lambda_{1,4}} + L_{-1} \ket{\phi_{1,2}} \qquad \text{and} \qquad L_1 \ket{\lambda_{1,4}} = \beta_{1,4} \ket{\phi_{1,2}}.
\end{equation}
The $\IndMod{1,4}$ vanishing singular vector lifts to a vanishing vector in $\LogMod{1,4}$ of the form
\begin{multline}
\ket{\xi_{1,4}} = \brac{L_{-4} - L_{-3} L_{-1} - L_{-2}^2 + \frac{5}{3} L_{-2} L_{-1}^2 - \frac{1}{4} L_{-1}^4} \ket{\lambda_{1,4}} \\
+ \brac{a_1 L_{-5} + a_2 L_{-4} L_{-1} + a_3 L_{-3} L_{-2} + a_4 L_{-2}^2 L_{-1}} \ket{\phi_{1,2}} = 0.
\end{multline}
Note that we are using the vanishing singular vector of $\IndMod{1,2}$ to replace $L_{-1}^2 \ket{\phi_{1,2}}$ by $\tfrac{2}{3} L_{-2} \ket{\phi_{1,2}}$, and thereby eliminate $3$ of the $7$ possible states descended from $\ket{\phi_{1,2}}$ at grade $5$.  Solving $L_1 \ket{\xi_{1,4}} = L_2 \ket{\xi_{1,4}} = 0$ then amounts to solving five linear equations in the five unknowns $a_1 , \ldots , a_4$ and $\beta_{1,4}$.  The unique solution is
\begin{equation}
a_1 = \frac{1}{2}, \qquad a_2 = \frac{4}{3}, \qquad a_3 = \frac{-8}{9}, \qquad a_4 = 0 \qquad \text{and} \qquad \beta_{1,4} = \frac{-1}{2}.
\end{equation}

This example is particularly easy (as is the analogous computation of $\beta_{1,5}$) because $\ket{\chi_{1,2}}$ (and $\ket{\chi_{1,1}}$) have such a simple form.  In general, computing in $\LogMod{1,s}$ requires choosing a gauge.  For instance, in $\LogMod{1,7}$ we have
\begin{equation} \label{eqnDefBeta17}
\ket{\chi_{1,5}} = \brac{L_{-3} - L_{-2} L_{-1} + \tfrac{1}{6} L_{-1}^3} \ket{\phi_{1,5}}, \qquad \text{hence} \qquad \beta_{1,7} = \bracket{\phi_{1,5}}{\brac{L_3 - L_1 L_2 + \tfrac{1}{6} L_1^3}}{\lambda_{1,7}}
\end{equation}
is the only gauge-invariant.  To fix the action of the positive Virasoro modes on $\ket{\lambda_{1,7}}$, it is convenient to work in the gauge in which $L_1 \ket{\lambda_{1,7}} = 0$.  We can choose such a gauge because there are three independent states at grade $3$ in the submodule $\IndMod{1,5}$, hence there are two effective\footnote{Recall that gauge transforms corresponding to shifts by the singular vector $\ket{\chi_{1,s'}}$ have no effect on $L_n \ket{\lambda_{1,s}}$ ($n>0$).} degrees of freedom in gauge transforming $\ket{\lambda_{1,7}}$, and $L_1 \ket{\lambda_{1,7}}$ belongs to the grade $2$ subspace of $\IndMod{1,5}$, which happens to be two-dimensional.  The invariance of \eqnref{eqnDefBeta17} then gives
\begin{equation}
L_2 \ket{\lambda_{1,7}} = \frac{-1}{8} \beta_{1,7} L_{-1} \ket{\phi_{1,5}} \qquad \text{and} \qquad L_3 \ket{\lambda_{1,7}} = \frac{1}{2} \beta_{1,7} \ket{\phi_{1,5}}
\end{equation}
in this gauge.

The vanishing singular vector of $\IndMod{1,7}$ is at grade $7$, so there exists a vanishing singular vector $\ket{\xi_{1,7}}$ in $\LogMod{1,7}$ at grade $10$.  Referring to the character of $\LogMod{1,7}$, one finds that $L_1 \ket{\xi_{1,7}} = L_2 \ket{\xi_{1,7}} = 0$ reduces to $62$ linear equations in $51$ unknowns (one of which is $\beta_{1,7}$).  Computing these equations in the gauge described above and solving them takes only a few seconds\footnote{We again used \textsc{Maple 10} for these calculations.}, giving a unique solution (as it must), with $\beta_{1,7} = \tfrac{-35}{3}$.  This matches the result obtained \cite{RidPer07} from the Nahm-Gaberdiel-Kausch algorithm (given in \eqnref{eqnOldBetas23}).

We outline one further example to illustrate a subtlety that one sometimes encounters when gauge-fixing.  To compute $\beta_{1,8}$, we note that it is obtained by acting on $\ket{\lambda_{1,8}}$ with a degree-$6$ composite Virasoro operator (whose exact form we will omit).  The \hwsm{} $\IndMod{1,4}$ has $9$ independent states at grade $6$, so there are $8$ independent effective gauge transformations that we can apply to $\ket{\lambda_{1,8}}$.  There are $6$ and $4$ independent states at grades $5$ and $4$ respectively, so we can choose the $8$ independent gauge transformations so as to tune all $6$ of the coefficients of $L_1 \ket{\lambda_{1,8}}$ to zero, and additionally, a further $2$ of the $4$ coefficients of $L_2 \ket{\lambda_{1,8}}$ to zero.  For definiteness, we will work in the gauge
\begin{equation} \label{eqnGauge18}
L_1 \ket{\lambda_{1,8}} = 0 \qquad \text{and} \qquad L_2 \ket{\lambda_{1,8}} = \brac{a L_{-4} + b L_{-2}^2} \ket{\phi_{1,4}}.
\end{equation}
Then,
\begin{align}
L_3 \ket{\lambda_{1,8}} &= - \brac{\brac{5a+3b} L_{-3} + 6b L_{-2} L_{-1}} \ket{\phi_{1,4}} & L_5 \ket{\lambda_{1,8}} &= -10 \brac{a+3b} L_{-1} \ket{\phi_{1,4}} \\
L_4 \ket{\lambda_{1,8}} &= \brac{\brac{10a+12b} L_{-2} + 9b L_{-1}^2} \ket{\phi_{1,4}} & L_6 \ket{\lambda_{1,8}} &= 5 \brac{a+3b} \ket{\phi_{1,4}},
\end{align}
from which we determine that $\beta_{1,8} = \tfrac{-112}{3} a - \tfrac{1774}{9} b$.  This is a strange result, as we should be able to fix $L_n \ket{\lambda_{1,8}}$ ($n>0$) in terms of the gauge-invariant $\beta_{1,8}$.  There are no other invariants, so we conclude that there must exist an additional relation between $a$ and $b$.

And indeed there is, though it is somewhat delicate to find:  Let $\ket{\zeta_{1,4}}$ denote the vanishing singular vector of the \hwsm{} $\IndMod{1,4}$.  Obviously, $\bracket{\zeta_{1,4}}{L_2}{\lambda_{1,8}} = 0$.  However, if we compute $L_{-2} \ket{\zeta_{1,4}}$ explicitly, take its adjoint, and apply it to $\ket{\lambda_{1,8}}$ with the above choice of gauge, we obtain
\begin{equation}
\bracket{\zeta_{1,4}}{L_2}{\lambda_{1,8}} = 126 a + 324 b.
\end{equation}
A similar computation using $L_1^2$ instead of $L_2$ yields $84 a + 216 b$, so in both cases we conclude that $b = \tfrac{-7}{18} a$.  The gauge-fixing is now complete, so we can compute $\beta_{1,8}$ as before.  This is somewhat more computationally intensive --- we must determine and solve $152$ linear equations in $116$ unknowns --- but an hour and a half of computation gives a unique solution with
\begin{equation}
\beta_{1,8} = \frac{-13475}{216}.
\end{equation}
We have confirmed this value for the logarithmic coupling via a (very tedious) application of the Nahm-Gaberdiel-Kausch algorithm to the fusion of $\IndMod{1,3}$ and $\IndMod{1,6}$ (to grade $6$).  We have also checked through explicit calculation that this value of $\beta_{1,8}$ is not dependent upon our gauge choice (\ref{eqnGauge18}).

\section{$\LogMinMod{2}{5}$:  The Yang-Lee Edge Singularity} \label{secLM25}

We now generalise the analysis of \secref{secLM23} to $\LogMinMod{2}{5}$.  Just as $\LogMinMod{2}{3}$ has been shown to describe crossing probabilities for critical percolation at $c=c_{2,3}=0$, we expect that the logarithmic theory we will construct below describes similar non-local observables in the Yang-Lee edge singularity at $c=c_{2,5}=\tfrac{-22}{5}$.  The thermodynamic limit of this statistical model has been previously identified (as a bulk \cft{}) with the minimal model $\MinMod{2}{5}$ \cite{CarCon85}.

As with $\LogMinMod{2}{3}$, we begin with the observation that the vacuum module in this theory cannot be irreducible.  In fact, none of the modules corresponding to the $\MinMod{2}{5}$ Kac table can be irreducible, for fusing such a module with itself would yield the irreducible vacuum module, and the vanishing of the vacuum singular vector at grade $4$ implies that the theory is $\MinMod{2}{5}$ (at least on the chiral level) \cite{FeiAnn92}.  The vacuum module of $\LogMinMod{2}{5}$ must therefore be again of the form $\IndMod{1,1}$ and we will \emph{assume} that the other modules corresponding to the Kac table are also present as $\IndMod{1,s}$, $s=2,3,4$.  Our proposal is therefore to modify $\MinMod{2}{5}$ by replacing the $4$ irreducible modules $\IrrMod{1,s}$, $1\leqslant s \leqslant 4$, by their reducible, but indecomposable, versions $\IndMod{1,s}$ (note that this modification breaks the Kac symmetry of $\MinMod{2}{5}$).  These indecomposable modules are the building blocks of our $\LogMinMod{2}{5}$ model.

\begin{table}
\begin{center}
\setlength{\extrarowheight}{4pt}
\begin{tabular}{|C|C|C|C|C|C|C|C|C|C|C}
\hline
0 & \tfrac{-1}{5} & \tfrac{-1}{5} & 0 & \tfrac{2}{5} & 1 & \tfrac{9}{5} & \tfrac{14}{5} & 4 & \tfrac{27}{5} & \cdots \\[1mm]
\hline
\tfrac{11}{8} & \tfrac{27}{40} & \tfrac{7}{40} & \tfrac{-1}{8} & \tfrac{-9}{40} & \tfrac{-1}{8} & \tfrac{7}{40} & \tfrac{27}{40} & \tfrac{11}{8} & \tfrac{91}{40} & \cdots \\[1mm]
\hline
4 & \tfrac{14}{5} & \tfrac{9}{5} & 1 & \tfrac{2}{5} & 0 & \tfrac{-1}{5} & \tfrac{-1}{5} & 0 & \tfrac{2}{5} & \cdots \\[1mm]
\hline
\end{tabular}
\vspace{3mm}
\caption{The first three rows of the extended Kac table for $c=c_{2,5}=\tfrac{-22}{5}$, listing the dimensions $h_{r,s}$ of the primary fields $\phi_{r,s}$.  Here, $r$ increases downwards, and $s$ increases to the right, so the top-left-hand corner corresponds to the identity field $\phi_{1,1}$, which with $\phi_{1,2}$, $\phi_{1,3}$ and $\phi_{1,4}$ exhausts the Kac table of $\MinMod{2}{5}$.} \label{tabExtKacc=-22/5}
\end{center}
\end{table}

The fusion of these modules can again be calculated using the Nahm-Gaberdiel-Kausch algorithm, out of which a picture of the logarithmic structure of $\LogMinMod{2}{5}$ emerges gradually.  In particular, $\IndMod{1,1}$ is again the identity of the fusion ring, and we find the non-trivial fusion rules
\begin{align} \label{eqnFusions25}
\IndMod{1,2} \fuse \IndMod{1,2} &= \IndMod{1,1} \oplus \IndMod{1,3} & \IndMod{1,2} \fuse \IndMod{1,3} &= \IndMod{1,2} \oplus \IndMod{1,4} \notag \\
\IndMod{1,2} \fuse \IndMod{1,4} &= \IndMod{1,3} \oplus \IndMod{1,5} & \IndMod{1,2} \fuse \IndMod{1,5} &= \LogMod{1,6} \notag \\
\IndMod{1,3} \fuse \IndMod{1,3} &= \IndMod{1,1} \oplus \IndMod{1,3} \oplus \IndMod{1,5} & \IndMod{1,3} \fuse \IndMod{1,4} &= \IndMod{1,2} \oplus \LogMod{1,6} \\
\IndMod{1,3} \fuse \IndMod{1,5} &= \IndMod{1,5} \oplus \LogMod{1,7} & \IndMod{1,4} \fuse \IndMod{1,4} &= \IndMod{1,1} \oplus \IndMod{1,5} \oplus \LogMod{1,7} \notag \\
\IndMod{1,4} \fuse \IndMod{1,5} &= \LogMod{1,6} \oplus \LogMod{1,8} & \IndMod{1,5} \fuse \IndMod{1,5} &= \IndMod{1,5} \oplus \LogMod{1,7} \oplus \LogMod{1,9}. \notag
\end{align}
As expected, we again generate rank $2$ staggered modules $\LogMod{1,s}$ (when $5 \nmid s$) on which $L_0$ is non-diagonalisable.  The structure of these modules is extremely similar to those described in the case of $\LogMinMod{2}{3}$:  They again have a maximal \hwsm{} isomorphic to $\IndMod{1,s'}$, but with $s'$ no longer given by \eqnref{eqnHWSMLabel23}, but rather by
\begin{equation} \label{eqnHWSMLabel25}
s' = 
\begin{cases}
s-2 & \text{if $s = 1 \pmod{5}$,} \\
s-4 & \text{if $s = 2 \pmod{5}$,} \\
s-6 & \text{if $s = 3 \pmod{5}$,} \\
s-8 & \text{if $s = 4 \pmod{5}$.}
\end{cases}
\end{equation}
Moreover, as before, the quotient of $\LogMod{1,s}$ by $\IndMod{1,s'}$ is isomorphic to the \hwm{} $\IndMod{1,s}$.  These staggered modules therefore give rise to a logarithmic structure for $\LogMinMod{2}{5}$ in exactly the same way that it arose in $\LogMinMod{2}{3}$.  The dimensions of the \hwss{} appearing in these modules are displayed in \tabref{tabExtKacc=-22/5}, which presents a part of the extended Kac table for $c=\tfrac{-22}{5}$.

We note that the fusion rules (\ref{eqnFusions25}) are in perfect agreement with the procedure given for computing the $\LogMinMod{2}{3}$ fusion rules in \secref{secLM23}, with $s'$ given as above.  We have further tested these rules with more general fusions involving staggered modules, for example
\begin{equation}
\IndMod{1,2} \fuse \LogMod{1,6} = 2 \IndMod{1,5} \oplus \LogMod{1,7}, \qquad \IndMod{1,3} \fuse \LogMod{1,6} = 2 \LogMod{1,6} \oplus \LogMod{1,8},
\end{equation}
again with perfect agreement.  We conclude that the spectrum of $\LogMinMod{2}{5}$ must contain
\begin{equation}
\set{\IndMod{1,s} \st 5 \mid s \geqslant 5} \cup \set{\LogMod{1,s} \st 5 \nmid s > 5}.
\end{equation}

Finally, we have computed the logarithmic couplings $\beta_{1,s}$ for the staggered modules appearing in the fusion rules (\ref{eqnFusions25}).  These are defined in the same way as in $\LogMinMod{2}{3}$ (\eqnref{eqnDefLogCoupling}):  The \hwsm{} is generated by the \hws{} $\ket{\phi_{1,s'}}$ which has a descendant non-vanishing singular vector $\ket{\chi_{1,s'}}$ with a Jordan partner state  $\ket{\lambda_{1,s}}$. $\beta_{1,s}$ is then defined to be $\braket{\chi_{1,s'}}{\lambda_{1,s}}$.  In each case, the coupling obtained from the Nahm-Gaberdiel-Kausch algorithm coincided precisely with that obtained directly by choosing a gauge and solving for the vanishing singular vector descended from $\ket{\lambda_{1,s}}$.  This is a strong confirmation of these results, which are
\begin{equation} \label{eqnBetas25}
\beta_{1,6} = \frac{-3}{2}, \qquad \beta_{1,7} = \frac{21}{8}, \qquad \beta_{1,8} = \frac{189}{8} \qquad \text{and} \qquad \beta_{1,9} = \frac{77}{8}.
\end{equation}

\section{General $\LogMinMod{2}{p}$ Theories} \label{secLM2p}

Our construction of $\LogMinMod{2}{3}$ and $\LogMinMod{2}{5}$ can easily be generalised to define a corresponding theory $\LogMinMod{2}{p}$, for every odd $p \geqslant 3$.  An identical argument to that given in \secref{secLM25} shows that none of the modules corresponding to the Kac table of $\MinMod{2}{3}$ can be irreducible, and our proposal is simply to replace each and every one of them with its reducible, but indecomposable, counterpart $\IndMod{1,s}$ ($s = 1 , \ldots , p-1$).  For the vacuum module, this is the only possibility; our assumption then is that the other modules corresponding to the Kac table are present \emph{and} have this particular indecomposable structure.

These modules $\IndMod{1,s}$, $s = 1 , \ldots , p-1$, then generate (see \secref{secInconsistency}) the spectrum of $\LogMinMod{2}{p}$ via fusion.  The fusion rules of $\LogMinMod{2}{p}$ are given\footnote{We have verified that this indeed holds for various examples with $p=7$ and $p=9$.} by the procedure described in \secref{secLM23} (just after \eqnref{eqnMFuseM}), except that $s'$ is defined by
\begin{equation}
s' = s - 2 \sigma, \qquad \text{where} \qquad s = \sigma \pmod{p} \quad \text{and} \quad \sigma \in \set{1 , 2 , \ldots , p-1}.
\end{equation}
The fusion of $\IndMod{1,2}$ and $\IndMod{1,p-1}$ therefore generates $\IndMod{1,p}$, and this in turn generates staggered modules $\LogMod{1,s}$:
\begin{equation}
\IndMod{1,2} \fuse \IndMod{1,p} = \LogMod{1,p+1}, \qquad \IndMod{1,3} \fuse \IndMod{1,p} = \IndMod{1,p} \oplus \LogMod{1,p+2}, \qquad \ldots
\end{equation}
The staggered modules appearing in our general $\LogMinMod{2}{p}$ models are identical in structure to those we have discussed above, and we will use the same notation to denote them.  In particular, the generating primary state will be denoted by $\ket{\phi_{1,s'}}$, its non-vanishing singular descendant by $\ket{\chi_{1,s'}}$, and the logarithmic partner of this descendant by $\ket{\lambda_{1,s}}$ (as in \figref{figStagMod}).

The $\LogMinMod{2}{p}$ models are therefore logarithmic \cfts{} whose spectrum must contain
\begin{equation} \label{eqnSpec2p}
\set{\IndMod{1,s} \st p \mid s \geqslant p} \cup \set{\LogMod{1,s} \st p \nmid s > p}.
\end{equation}
These generalise the chiral parts of the minimal models $\MinMod{2}{p}$ in two obvious ways:  They share the same central charge, and their spectra are generated by replacing every irreducible module $\IrrMod{1,s}$ of the (chiral) minimal model by its indecomposable counterpart $\IndMod{1,s}$.  Since these theories follow minimally from the single assumption that \emph{every} (chiral) primary field of $\MinMod{2}{p}$ has a primary counterpart in $\LogMinMod{2}{p}$ (with a single non-vanishing singular vector), we believe that it is natural to refer to these theories as \emph{logarithmic chiral minimal models}\footnote{We will see in \secref{secDual} that logarithmic theories may be constructed from other modules, in particular from the modules appearing in the first column of the extended Kac table.  These theories do not share this property.  Whilst they have the correct central charge, they do not possess modules naturally corresponding to those of the minimal model.}.  Our expectation (albeit vague but borne out for $p=3$), that they describe non-local observables in the statistical models whose local observables are described by $\MinMod{2}{p}$, gives another reason to single out these theories as natural extensions of these minimal models.  We present, for later reference, a portion of the extended Kac table for $c = c_{2,p} = 1 - 3 \brac{p-2}^2 / p$ in \tabref{tabExtKacc=c2p}.

\begin{table}
\begin{center}
\setlength{\extrarowheight}{4pt}
\begin{tabular}{C||C|C|C|C|C|C|C|C|C|C|C|C|C}
\raisebox{-0.7ex}{r} \mspace{-3mu} \diagdown \mspace{-4mu} \raisebox{0.7ex}{s} & 1 & 2 & \cdots & p-2 & p-1 & p & p+1 & p+2 & \cdots & 2p-1 & 2p & 2p+1 & \cdots \\[1mm]
\hline \hline
1 & 0 & \tfrac{3-p}{2p} & \cdots & \tfrac{3-p}{2p} & 0 & \tfrac{p-1}{2p} & 1 & \tfrac{3 \brac{p+1}}{2p} & \cdots & p-1 & & p+2 & \cdots \\[1mm]
\hline
2 & \tfrac{3p-4}{8} & & \cdots & & \tfrac{4-p}{8} & & & & \cdots & & & & \cdots \\[1mm]
\hline
3 & p-1 & & \cdots & \tfrac{3 \brac{p+1}}{2p} & 1 & \tfrac{p-1}{2p} & 0 & \tfrac{3-p}{2p} & \cdots & 0 & & 1 & \cdots \\[1mm]
\hline
\end{tabular}
\vspace{3mm}
\caption{A part of the extended Kac table for $c=c_{2,p}$, listing (some of) the dimensions $h_{r,s}$ of the primary fields $\phi_{r,s}$.  The Kac table of $\MinMod{2}{p}$ corresponds to the $h_{1,s}$ with $1 \leqslant s \leqslant p-1$.} \label{tabExtKacc=c2p}
\end{center}
\end{table}

In order to be able to calculate within these logarithmic minimal models, one needs to compute the logarithmic couplings $\beta_{1,s} = \braket{\chi_{1,s'}}{\lambda_{1,s}}$.  As we have seen, there are at least two ways to do this, though both become computationally prohibitive as the grade of the singular vectors of the modules increases.  Surprisingly however, it is possible to deduce general formulae for certain logarithmic couplings using the Nahm-Gaberdiel-Kausch algorithm.

The idea behind these deductions is not subtle, though it does require some knowledge of how this algorithm works.  We refer to \cite{GabInd96} for a thorough account of the application of this algorithm.  What is relevant is that the structure of the module $\LogMod{}$ that results from fusing two $\alg{Vir}$-modules, $\IndMod{}$ and $\IndMod{}'$, can be analysed through calculations in a finite-dimensional vector space.  Specifically, if we want to analyse the structure to grade $n$, then we compute within a space whose dimension is given by the grade of the first vanishing singular vector in $\IndMod{}$, multiplied by the dimension of the subspace of $\IndMod{}'$ consisting of states of grade less than or equal to $n$.  A vital observation is that often this \emph{working space} is in fact strictly larger than the actual subspace of $\LogMod{}$ of grade $\leqslant n$ states which we are trying to analyse.  In this situation, one has to reduce the dimensionality of the working space by determining a so-called \emph{spurious subspace} \cite{NahQua94}, before the analysis proper can begin.  This determination is often the most computationally intensive part of the algorithm.

What we want to do is compute the fusion rule
\begin{equation} \label{eqnFusn+1byp}
\IndMod{1,n+1} \fuse \IndMod{1,p} = 
\begin{cases}
\LogMod{1,p+n} \oplus \LogMod{1,p+n-2} \oplus \ldots \oplus \LogMod{1,p+1} & \text{if $n$ is odd,} \\
\LogMod{1,p+n} \oplus \LogMod{1,p+n-2} \oplus \ldots \oplus \LogMod{1,p+2} \oplus \IndMod{1,p} & \text{if $n$ is even.}
\end{cases}
\end{equation}
to grade $n$.  Since the dimensions of the generating states of $\LogMod{1,p+n}$ differ by $h_{1,p+n} - h_{1,p-n} = n$, knowledge of the explicit form of the non-vanishing singular vector of $\IndMod{1,p-n} \subset \LogMod{1,p+n}$ (which is at grade $n$) will then allow us to compute $\beta_{1,p+n}$.  For small $n$, this is indeed feasible, \emph{provided} that we can apply the Nahm-Gaberdiel-Kausch algorithm to this somewhat general fusion rule.

In this application, we require the explicit form of the vanishing singular vector of the first module appearing in the fusion:  $\IndMod{1,n+1}$.  This is at grade $n+1$, so again, for small $n$, this can be determined explicitly (as a function of $p$).  The vanishing singular vector of the second module, $\IndMod{1,p}$, is not so easily computed (for general $p$), but it enters into the fusion algorithm in only two ways.  First, it is used to derive a basis for $\IndMod{1,p}$ to grade $n$.  But if $p > n$, then this singular vector will be too deep to affect the basis.  Second, it is used in an essential way to determine the spurious subspace.  But note that (for $p>n$) the dimension of the working space is $\brac{n+1} \sum_{m=0}^n \partnum{m}$, where $\partnum{m}$ denotes the number of partitions of $m$.  If $p > 2n$, then the vanishing singular vectors of every module appearing on the right hand side of \eqnref{eqnFusn+1byp} occur at grades greater than $n$ (the lowest-graded one is that of $\IndMod{1,p-n} \subset \LogMod{1,p+n}$ at grade $p-n$), so the dimension of the subspace of the right hand side consisting of states of grade $\leqslant n$ is also $\brac{n+1} \sum_{m=0}^n \partnum{m}$.  There is therefore no spurious subspace, and hence no need to compute the singular vector of $\IndMod{1,p}$ explicitly.

This argument therefore allows us to compute (in principle) an explicit expression for $\beta_{1,p+n}$ for all $p > 2n$.  In practise, computational limitations restrict us to small $n$ only\footnote{In fact, for $n > 2$, it is computationally worthwhile to consider, instead of (\ref{eqnFusn+1byp}), the fusion rule
\begin{equation*}
\IndMod{1,2} \fuse \LogMod{1,p+n-1} = \LogMod{1,p+n} \oplus \LogMod{1,p+n-2} \qquad \text{($n > 2$).}
\end{equation*}
The analysis is identical, except that it is necessary to choose a gauge in order to fix the Virasoro action on $\LogMod{1,p+n-1}$.}.  We have computed
\begin{align}
\beta_{1,p+1} &= - \frac{p-2}{2} & \text{($p > 2$)}, \label{eqnGenBeta1,p+1} \\
\beta_{1,p+2} &= \frac{\brac{p-4} \brac{p-2} \brac{p+2}}{8} & \text{($p > 4$)}, \label{eqnGenBeta1,p+2} \\
\beta_{1,p+3} &= - \frac{\brac{p-6} \brac{p-2} \brac{p+2} \brac{p+4}}{8 \brac{p-4}} & \text{($p > 6$)}, \label{eqnGenBeta1,p+3} \\
\beta_{1,p+4} &= \frac{\brac{p-8} \brac{p-6} \brac{p-4} \brac{p-2} \brac{p+2} \brac{p+4} \brac{p+6}}{8 \brac{p^2-8p+24}^2} & \text{($p > 8$)}, \label{eqnGenBeta1,p+4}
\end{align}
though we note that these formulae also correctly predict
\begin{equation}
\beta_{1,5} = \frac{-5}{8} \quad \text{when $p=3$} \qquad \text{and} \qquad \beta_{1,8} = \frac{189}{8}, \quad \beta_{1,9} = \frac{77}{8} \quad \text{when $p=5$}
\end{equation}
(refer to \eqnDref{eqnOldBetas23}{eqnBetas25}).  We have also confirmed these formulae in several computations with $p=7$ and $p=9$, again even for $p$ outside the ranges specified.  In particular, we have (arduously) computed from singular vectors that $\beta_{1,11} = \tfrac{-19305}{2312}$ for $p=7$, in full agreement with \eqnref{eqnGenBeta1,p+4}.

It is rather striking that \eqnsref{eqnGenBeta1,p+1}{eqnGenBeta1,p+4} involve such a simple series of linear factors in $p$, particularly upon recalling that the precise values of these logarithmic couplings depend upon the (arbitrary) normalisation of the singular vector $\ket{\chi_{1,p-n}}$ chosen in \eqnref{eqnSVNorm}.  This observation is given more weight upon noting that the factors of $p-4$ and $p^2-8p+24$ which appear in the denominators of \eqnDref{eqnGenBeta1,p+3}{eqnGenBeta1,p+4} (respectively) also appear in the explicit expressions for the normalised $\ket{\chi_{1,p-n}}$.  What this suggests is that there is in fact a more natural normalisation for these singular vectors.  Indeed, this (admittedly small) set of data is consistent with the conjecture that upon normalising this non-vanishing singular vector such that\footnote{In hindsight, it is perhaps not so surprising that normalising the coefficient of $L_{-1}^n$ is more natural than normalising that of $L_{-n}$.  The latter coefficient is sensitive to how we choose to order Virasoro modes --- the usual Poincar\'{e}-Birkhoff-Witt order is by no means canonical --- whereas the former is not!} (compare with \eqnref{eqnSVNorm})
\begin{equation} \label{eqnRenormalisedSV}
\ket{\chi_{1,p-n}} = \brac{\ldots - \frac{p^n}{2 \brac{n-1}!} L_{-1}^n} \ket{\phi_{1,p-n}},
\end{equation}
then the corresponding \emph{renormalised} logarithmic couplings take the form
\begin{equation} \label{eqnBeta1,p+n}
\hat{\beta}_{1,p+n} = \brac{-1}^n \frac{p}{8} \prod_{i=-n}^{n-1} \brac{p+2i}.
\end{equation}
We expect that such a formula will hold for all $p > n$.  Note that this normalisation differs from that given in the general (combinatorial) formula of \cite{BenDeg88} by a factor of $-\tfrac{1}{2} n! p^n$.

There is one further logarithmic coupling that we have been able to compute in general:
\begin{equation}
\beta_{1,2p+1} = -\frac{3 \brac{p-2} \brac{3p-2} \brac{p+1} \brac{p+2}}{2 p^2 \brac{p-1}}.
\end{equation}
This correctly gives $\beta_{1,7} = \tfrac{-35}{3}$ for $p=3$ (\eqnref{eqnOldBetas23}), and predicts $\beta_{1,11} = \tfrac{-2457}{100}$ for $p=5$.  Half an hour of computation confirms this latter value directly from the singular vectors.  We remark that if we renormalise the non-vanishing singular vector $\ket{\chi_{1,2p-1}}$ so that the coefficient of $L_{-1}^3$ is $-p/2$, then the renormalised logarithmic coupling takes the form
\begin{equation}
\hat{\beta}_{1,2p+1} = -\frac{3p}{8} \brac{3p-2} \prod_{i=-2}^{2} \brac{p+i}.
\end{equation}
This again is strikingly simple.

\section{Inconsistencies in Logarithmic Theories} \label{secInconsistency}

In principle then, the considerations of the previous sections allow us to compute within a general staggered module, hence within $\LogMinMod{2}{p}$ (or at least the part corresponding to the first row of the extended Kac table).  A natural question to ask now is whether there are any further modules present in the theory.  In this section, we will show that augmenting $\LogMinMod{2}{p}$ by such modules is fraught with difficulty, by proving that including the most obvious candidate modules in the spectrum leads to fundamental contradictions with the requirements of conformal invariance.

Let us begin with $\LogMinMod{2}{3}$.  As we have seen, the module $\IndMod{1,2}$ generates the set of modules (\ref{eqnSpec23}), each of which can be associated with an entry in the first row of the extended Kac table.  We will first attempt to add the module $\IndMod{2,1}$ to the spectrum.  In \cite{RidPer07}, we showed that fusing this module with itself generates a rank $2$ staggered module $\LogMod{3,1}$, whose presence in the theory is inconsistent with the presence of $\LogMod{1,5}$.  This was shown using an argument of Gurarie and Ludwig \cite[App.\ A]{GurCon04}.  Rather than present this case again, we will instead detail a slightly more involved argument (whose conclusion is identical).  The advantage of this latter argument is that it generalises directly to all $p$.

We therefore compute with the Nahm-Gaberdiel-Kausch algorithm (see also \cite{EbeVir06,RasFus07}) that fusing $\IndMod{1,2}$ with $\IndMod{2,1}$ gives $\IndMod{2,2}$.  Fusing this latter module again with $\IndMod{2,1}$ now yields a rank $2$ staggered module which we will denote by $\LogMod{3,2}$.  Its \hwsm{} is isomorphic to $\IndMod{1,2}$ and the quotient by this submodule is isomorphic to $\IndMod{3,2}$.  As $h_{3,2} = h_{1,4} = 1$ (\tabref{tabExtKacc=0}), we see that $\LogMod{3,2}$ is structurally very similar to $\LogMod{1,4}$.  Indeed, the only difference is that $\ket{\lambda_{1,4}}$ has a vanishing descendant at grade $4$ whereas $\ket{\lambda_{3,2}}$ has no vanishing descendant until grade $6$.  This difference in vanishing singular vectors leads to different logarithmic couplings, $\beta_{1,4} = \tfrac{-1}{2}$ whereas $\beta_{3,2} = \tfrac{1}{3}$.

This is where we generalise Gurarie and Ludwig's argument.  The \opes{} that result from the structures of $\LogMod{1,4}$ and $\LogMod{3,2}$ are (we use the standard convention $z_{12} = z_1 - z_2$ for brevity)
\begin{equation} \label{eqnOPETlam}
\func{T}{z_1} \func{\lambda_{r,s}}{z_2} = \frac{\beta_{r,s} \func{\phi_{1,2}}{z_2}}{z_{12}^3} + \frac{\func{\lambda_{r,s}}{z_2} + \func{\chi_{1,2}}{z_2}}{z_{12}^2} + \frac{\partial \func{\lambda_{r,s}}{z_2}}{z_{12}} + \ldots \qquad \text{($\brac{r,s} = \brac{3,2} , \brac{1,4}$).}
\end{equation}
We will derive a contradiction by using these expansions to compute the $2$-point function $\corrfn{\func{\lambda_{1,4}}{z_1} \func{\lambda_{3,2}}{z_2}}$.  To do this however, we first need to compute a few auxiliary results.  We begin with $\corrfn{\func{\chi_{1,2}}{z_1} \func{\lambda_{3,2}}{z_2}}$, noting that $\chi_{1,2} = \partial \phi_{1,2}$.  As usual, the global conformal invariance generated by $L_{-1}$ and $L_0$ (the fact that both $\ket{0}$ and $\bra{0}$ are annihilated by these operators) leads to an \ode{} for this correlator:
\begin{equation} \label{eqnODEchilam}
\brac{z_{12} \diff{}{z_{12}} + 2} \corrfn{\func{\chi_{1,2}}{z_1} \func{\lambda_{3,2}}{z_2}} = -\corrfn{\func{\chi_{1,2}}{z_1} \func{\chi_{1,2}}{z_2}} = 0.
\end{equation}
The solution is then (see \appref{appFieldAdjoints} for the sign appearing here)
\begin{equation} \label{eqnCF23:12,32}
\corrfn{\func{\chi_{1,2}}{z_1} \func{\lambda_{3,2}}{z_2}} = \frac{-\beta_{3,2}}{z_{12}^2}.
\end{equation}
Similarly, $\corrfn{\func{\chi_{1,2}}{z_1} \func{\lambda_{1,4}}{z_2}} = -\beta_{1,4} / z_{12}^2$, so
\begin{equation} \label{eqnCF23:14,12}
\corrfn{\func{\lambda_{1,4}}{z_1} \func{\chi_{1,2}}{z_2}} = \frac{-\beta_{1,4}}{z_{12}^2},
\end{equation}
since $\lambda_{1,4}$ and $\chi_{1,2}$ are mutually bosonic (\appref{appMutualBosons}).

We are now ready to tackle $\corrfn{\func{\lambda_{1,4}}{z_1} \func{\lambda_{3,2}}{z_2}}$.  The \opes{} (\ref{eqnOPETlam}) and global conformal invariance under $L_{-1}$ and $L_0$ induce the \ode{}
\begin{equation}
\brac{z_{12} \diff{}{z_{12}} + 2} \corrfn{\func{\lambda_{1,4}}{z_1} \func{\lambda_{3,2}}{z_2}} = -\corrfn{\func{\chi_{1,2}}{z_1} \func{\lambda_{3,2}}{z_2}} - \corrfn{\func{\lambda_{1,4}}{z_1} \func{\chi_{1,2}}{z_2}} = \frac{\beta_{3,2} + \beta_{1,4}}{z_{12}^2},
\end{equation}
the solution of which is
\begin{equation} \label{eqnCorrFn23:14,32}
\corrfn{\func{\lambda_{1,4}}{z_1} \func{\lambda_{3,2}}{z_2}} = \frac{C + \brac{\beta_{3,2} + \beta_{1,4}} \log z_{12}}{z_{12}^2},
\end{equation}
for some unknown\footnote{In fact, $C$ is not just unknown, but \emph{unknowable}.  More specifically, it is not gauge-invariant.  Shifting either of the fields appearing in this correlation function by multiples of $\chi_{1,2}$ leads to shifts in the constant $C$ by multiples of the (corresponding) logarithmic coupling.  We mention that logarithms in correlators are always accompanied by unknowable constants.  This is required so that some semblance of locality is preserved:  Swapping $z_1$ and $z_2$ leads to ill-defined constants of the form $\log \brac{-1}$ in such correlators, which can then be absorbed by the unknowable constant $C$.  Similarly, logarithms in these functions request the presence of a dimensionful scale (breaking conformal invariance!), and this too can be absorbed by the unknowable constant $C$.  In a logarithmic \cft{} then, locality and scale-invariance are only broken up to gauge transformations.  In particular, such a scale would not be physical.} constant $C$.  However, the global conformal invariance generated by $L_1$ induces another equation, a \pde{}, which must also be satisfied:
\begin{align}
\brac{z_1^2 \pardiff{}{z_1} + z_2^2 \pardiff{}{z_2} + 2 \brac{z_1 + z_2}} \corrfn{\func{\lambda_{1,4}}{z_1} \func{\lambda_{3,2}}{z_2}} &= -2 z_1 \corrfn{\func{\chi_{1,2}}{z_1} \func{\lambda_{3,2}}{z_2}} - 2 z_2 \corrfn{\func{\lambda_{1,4}}{z_1} \func{\chi_{1,2}}{z_2}} \notag \\
& \mspace{25mu} - \beta_{1,4} \corrfn{\func{\phi_{1,2}}{z_1} \func{\lambda_{3,2}}{z_2}} - \beta_{3,2} \corrfn{\func{\lambda_{1,4}}{z_1} \func{\phi_{1,2}}{z_2}} \notag \\
&= \frac{2 \brac{\beta_{3,2} z_1 + \beta_{1,4} z_2}}{z_{12}^2}.
\end{align}
It is not hard to verify that (\ref{eqnCorrFn23:14,32}) is a solution of this \pde{} \emph{if and only if} $\beta_{1,4} = \beta_{3,2}$.  This is a contradiction (they are not equal), so the conclusion is that the conformal invariance of the vacuum forbids simultaneously having $\LogMod{3,2}$ and $\LogMod{1,4}$ in the spectrum.  Backtracking, we conclude that $\IndMod{1,2}$ and $\IndMod{2,1}$ cannot both be in the spectrum either.

One can then ask if it is possible to add $\IndMod{2,s}$ to the spectrum for any $s$.  For $s=1,2$, we already know that the answer is ``no''.  $s=3$ is a little more interesting, as
\begin{equation}
\IndMod{1,2} \fuse \IndMod{2,3} = \JorMod{2,4},
\end{equation}
where $\JorMod{r,s}$ denotes a rank $2$ \emph{Jordan \hwm{}}\footnote{A rank $2$ Jordan \hwm{} is just a degenerate case of a rank $2$ staggered module in which the non-vanishing singular vector $\ket{\chi_{r',s'}}$ (in \figref{figStagMod}) coincides with the \hws{} $\ket{\phi_{r',s'}}$.} \cite{RohRed96}.  But, this $\JorMod{2,4}$ has \hwsm{} isomorphic to $\IndMod{2,2}$, and we have already seen that this is not allowed in the spectrum.  Continuing in this way, it is not difficult to show inductively (assuming the obvious fusion rules) that $\IndMod{2,s}$ cannot be consistently added to the spectrum for any $s \in \ZZ_+$.

One can take this further.  If we try to add $\IndMod{3,1}$ to the spectrum, then fusing it with $\IndMod{1,2}$ gives $\IndMod{3,2}$, and fusing $\IndMod{3,1}$ and $\IndMod{3,2}$ generates $\LogMod{3,2}$.  $\IndMod{3,1}$ therefore cannot be included in the spectrum.  In fact, this argument appears to extend to $\IndMod{r,1}$ for every $r > 1$, so we cannot include any of these modules.  Furthermore, by repeatedly fusing with $\IndMod{1,2}$ as above, it is easy to see that this means that \emph{every} $\IndMod{r,s}$ with $r > 1$ must be excluded from the spectrum.

Summarising, we see that the presence of $\IndMod{1,2}$ (and hence $\LogMod{1,4}$) in the spectrum of $\LogMinMod{2}{3}$ prevents any module of the form $\IndMod{r,s}$ with $r>1$ from appearing.  We have not ruled out augmentations by the more complicated staggered modules (of rank $2$ or higher), but we mention that such modules have \hwsms{} of the form $\IndMod{r,s}$, so these potential augmentations will also be severely constrained by the above arguments.  We conjecture then that the spectrum of this theory is precisely that given in (\ref{eqnSpec23}), which corresponds to the first row of the extended Kac table.

This argument can be generalised to all $\LogMinMod{2}{p}$.  Let us therefore attempt to add the module $\IndMod{2,1}$ to the spectrum (\ref{eqnSpec2p}).  Then, fusing repeatedly with $\IndMod{1,2}$, we generate $\IndMod{2,p-1}$.  By now, it should not be surprising (see \tabref{tabExtKacc=c2p}) that fusing $\IndMod{2,1}$ and $\IndMod{2,p-1}$ yields a staggered module $\LogMod{3,p-1}$ whose \hwsm{} is isomorphic to $\IndMod{1,p-1}$ and whose quotient by this submodule is isomorphic to $\IndMod{3,p-1}$.  This staggered module is therefore structurally identical to $\LogMod{1,p+1}$, except for having a vanishing logarithmic singular vector at grade $3 \brac{p-1} + 1 = 3p-2$ instead of at grade $p+2$ (here we measure grade with respect to the \hws{}).  We have already determined the logarithmic coupling $\beta_{1,p+1}$ in \eqnref{eqnGenBeta1,p+1}, and it is easy to determine $\beta_{3,p-1}$ in the same way.  The result is
\begin{equation} \label{eqnBetasNotEqual}
\beta_{3,p-1} = \frac{p-2}{p} \neq -\frac{p-2}{2} = \beta_{1,p+1}.
\end{equation}

The argument of Gurarie and Ludwig can now be generalised to our present situation.  Using the same procedure as before, we can compute the correlator $\corrfn{\func{\lambda_{1,p+1}}{z_1} \func{\lambda_{3,p-1}}{z_2}}$ by solving the \ode{} induced by the global conformal invariance generated by $L_{-1}$ and $L_0$ (up to an unknowable constant of integration).  Again, we find that this solution does not satisfy the \pde{} induced by $L_1$-global conformal invariance unless $\beta_{1,p+1} = \beta_{3,p-1}$, in contradiction with \eqnref{eqnBetasNotEqual}.  We therefore see that the presence of $\IndMod{1,2}$ in the spectrum of $\LogMinMod{2}{p}$ (for all odd $p$) is inconsistent with the presence of $\IndMod{2,p-1}$ and hence $\IndMod{2,1}$.

The computations which detail this argument are in fact identical to those which we have presented for $p=3$.  As with this previous case, it is easy to extend this argument to show that $\IndMod{1,2}$ and $\IndMod{r,s}$ are inconsistent for all $r > 1$.  Rather than present these arguments in repetitive detail, we would like to outline a more abstract (and less computational) argument, whose conclusions are nevertheless the same.  We want to emphasise however that the following argument is not entirely rigorous, and does not (at least in its present form) supersede that presented earlier.  Rather, we present it as an aid to understanding this somewhat subtle situation.

Suppose then that we had two staggered modules $\LogMod{1}$ and $\LogMod{2}$ such that the two logarithmic generators $\ket{\lambda_1}$ and $\ket{\lambda_2}$ (respectively) are both partners of dimension $h$ to the same non-vanishing singular vector $\ket{\chi}$ (for example, $\ket{\lambda_{1,2p-1}} \in \LogMod{1,2p-1}$ and $\ket{\lambda_{3,1}} \in \LogMod{3,1}$ both couple to the non-vanishing singular vector in the vacuum module $\IndMod{1,1}$).  Then, we have
\begin{equation}
\begin{split}
\brac{L_0 - h \id} \ket{\chi} &= 0 \\
\brac{L_0 - h \id} \ket{\lambda_1} &= \ket{\chi}
\end{split}
\quad \text{in $\LogMod{1}$,} \qquad \text{and} \quad
\begin{split}
\brac{L_0 - h \id} \ket{\chi} &= 0 \\
\brac{L_0 - h \id} \ket{\lambda_2} &= \ket{\chi}
\end{split}
\quad \text{in $\LogMod{2}$.}
\end{equation}
In each $\LogMod{i}$, a formal adjoint of $L_0 - h \id$ may be defined (with respect to a basis extending $\set{\ket{\chi} , \ket{\lambda_i}}$) \cite{RohRed96}.  This is a seemingly well-defined operator, but it must satisfy
\begin{equation}
\begin{split}
\brac{L_0 - h \id}^{\dagger} \ket{\chi} &= \ket{\lambda_1} \\
\brac{L_0 - h \id}^{\dagger} \ket{\lambda_1} &= 0
\end{split}
\quad \text{in $\LogMod{1}$,} \qquad \text{and} \quad
\begin{split}
\brac{L_0 - h \id}^{\dagger} \ket{\chi} &= \ket{\lambda_2} \\
\brac{L_0 - h \id}^{\dagger} \ket{\lambda_2} &= 0
\end{split}
\quad \text{in $\LogMod{2}$.}
\end{equation}
Since there is only one $\ket{\chi}$, this is a contradiction.

More precisely, what this argument suggests is that there cannot exist a module which has two submodules of the form $\LogMod{1}$ and $\LogMod{2}$.  But, this is exactly the structure that na\"{\i}vely extending our $\LogMinMod{2}{p}$ models by $\IndMod{r,s}$ ($r>1$) leads to.  We therefore conclude once again that such extensions are inconsistent, reinforcing our conjecture that the spectrum of $\LogMinMod{2}{p}$ is as given in (\ref{eqnSpec2p}) (corresponding to the first row of the extended Kac table).

We conclude this section with a remark.  This inconsistency result hinges on the fact that the logarithmic couplings of $\LogMod{1,p+1}$ and $\LogMod{3,p-1}$ can be explicitly computed, and are different (\eqnref{eqnBetasNotEqual}).  It is intriguing to observe that there are further pairs of couplings which can be naturally compared.  In particular, $\LogMod{1,p+n}$ and $\LogMod{3,p-n}$ (the latter decomposes as a vector space into $\IndMod{1,p-n}$ and $\IndMod{3,p-n}$) share the same \hwsm{}, hence the same non-vanishing singular vector.  We find that the logarithmic couplings of these modules are simply related by
\begin{equation} \label{eqnBetaRelation}
\beta_{3,p-n} = \frac{-2n}{p} \beta_{1,p+n},
\end{equation}
at least for $n \leqslant 4$ (for which explicit computations are possible, as in \eqnsref{eqnGenBeta1,p+1}{eqnGenBeta1,p+4}).  We conjecture that this relation continues to hold for all $n < p$.  Note that it is independent of the singular vector normalisation used to define the logarithmic couplings.

\section{Dual Theories $\DualLogMinMod{2}{p}$} \label{secDual}

We have just proven in \secref{secInconsistency} that the logarithmic \cfts{} we have constructed from $\IndMod{1,2}$ can not admit a module of the form $\IndMod{2,1}$ in their spectra.  It is therefore appropriate (and interesting) to ask the dual question:  Can we construct consistent logarithmic theories from $\IndMod{2,1}$, which (necessarily) will not admit a module of the form $\IndMod{1,2}$ in their spectra?  We would expect such a theory to in fact admit no module except for those corresponding to the first \emph{column} of the appropriate extended Kac table.  In other words, such a theory would be generated by $\IndMod{2,1}$, just as the $\LogMinMod{2}{p}$ are generated by $\IndMod{1,2}$.  It turns out that these \emph{dual} logarithmic models can indeed be constructed in this way, and we will denote them by $\DualLogMinMod{2}{p}$.

It should be stressed that $\DualLogMinMod{2}{p}$ is distinct from $\LogMinMod{2}{p}$, despite sharing the same central charge.  However, we have described the latter as a natural generalisation of the chiral minimal model $\MinMod{2}{p}$, in that it includes modules which clearly generalise those appearing in the Kac table of the minimal model.  This naturality property is lost with the dual theory $\DualLogMinMod{2}{p}$ --- they do not define a covering theory for the (chiral) minimal model $\MinMod{2}{p}$.

Nevertheless, this loss of naturality is no obstacle to being physically relevant.  In particular, we suspect that the dual model $\DualLogMinMod{2}{3}$ describes the scaling limit of the lattice model of \emph{self-avoiding walks} (which are supposed to model long polymer chains), just as $\LogMinMod{2}{3}$ describes the scaling limit of critical percolation.

The structures of the modules generated by fusing $\IndMod{2,1}$ with itself (repeatedly) can be investigated using the Nahm-Gaberdiel-Kausch algorithm.  Unsurprisingly, the results are quite similar to those we have described for $\LogMinMod{2}{p}$, and are perhaps even simpler.  We generate rank $2$ staggered modules $\LogMod{r,1}$, $r=3,5,7,\ldots$, which decompose as vector spaces into $\IndMod{r-2,1}$ and $\IndMod{r,1}$.  The spectrum of $\DualLogMinMod{2}{p}$ is therefore
\begin{equation}
\set{\IndMod{r,1} \st 2 \mid r \geqslant 2} \cup \set{\LogMod{r,1} \st 2 \nmid r > 2}.
\end{equation}
The fusion rules are conveniently described as follows:
\begin{enumerate}
\item Replace any $\LogMod{r,1}$ by the direct sum $\IndMod{r-2,1} \oplus \IndMod{r,1}$.
\item Compute the ``fusion'' using distributivity and the auxiliary rule
\begin{equation}
\IndMod{q,1} \fuse \IndMod{r,1} = \IndMod{\abs{q-r}+1,1} \oplus \IndMod{\abs{q-r}+3,1} \oplus \ldots \oplus \IndMod{q+r-3,1} \oplus \IndMod{q+r-1,1}.
\end{equation}
\item In the result, replace all direct sums of the form $\IndMod{r-2,1} \oplus \IndMod{r,1}$ by $\LogMod{r,1}$ (there is only one way to consistently do this).
\end{enumerate}

The logarithmic couplings of the staggered modules of these theories can be computed from the appropriate vanishing singular vectors, or using the Nahm-Gaberdiel-Kausch algorithm.  We have explicitly determined several $\DualLogMinMod{2}{p}$ couplings $\beta_{3,1}$ for various $p$ using the singular vector approach, and list them in \tabref{tabDualBetas} (the only other logarithmic coupling we have determined is $\beta_{5,1} = \frac{67375}{676}$ for $p=3$).  In contrast to $\LogMinMod{2}{p}$, we are not able to explicitly compute these couplings for general $p$:  computing $\beta_{3,1}$ for each $p$  requires being able to apply the Nahm-Gaberdiel-Kausch algorithm to grade $p-1$.

\begin{table}
\begin{center}
\setlength{\extrarowheight}{4pt}
\begin{tabular}{|C||C|C|C|C|}
\hline
p & 3 & 5 & 7 & 9 \\[1mm]
\hline
\beta_{3,1} & \frac{5}{6} & \frac{-77}{5} & \frac{984555}{5054} & \frac{-74364290}{173889} \\[1mm]
\hline
\hat{\beta}_{3,1} & \frac{15}{2} & -31185 & \frac{1550674125}{2} & -66409170077250 \\[1mm]
\hline
\end{tabular}
\vspace{3mm}
\caption{The logarithmic coupling $\beta_{3,1}$ (and its appropriately renormalised counterpart $\hat{\beta}_{3,1}$) for the staggered module $\LogMod{3,1}$ appearing in the $\DualLogMinMod{2}{p}$ theories.} \label{tabDualBetas}
\end{center}
\end{table}

However, we can \emph{predict} such a general formula for $\beta_{3,1}$, or rather for the \emph{renormalised} logarithmic coupling $\hat{\beta}_{3,1}$ which corresponds to normalising the non-vanishing singular vector $\ket{\chi_{1,1}}$ as in \eqnref{eqnRenormalisedSV} (with $n=p-1$).  First, note that \eqnref{eqnBeta1,p+n} conjectures a formula for the logarithmic coupling $\hat{\beta}_{1,2p-1}$ of $\LogMinMod{2}{p}$ (also renormalised according to \eqnref{eqnRenormalisedSV}):
\begin{equation}
\hat{\beta}_{1,2p-1} = \brac{-1}^{p-1} \frac{p}{8} \prod_{i=-p+1}^{p-2} \brac{p+2i} = \brac{-1}^{\brac{p-1} / 2} \frac{p!! \brac{3p-4}!!}{8},
\end{equation}
where $n!! = n \brac{n-2} \brac{n-4} \cdots 1$ (note that $p$ and $3p-4$ are odd).  \eqnref{eqnBetaRelation} now conjectures that
\begin{equation}
\hat{\beta}_{3,1} = \frac{-2 \brac{p-1}}{p} \hat{\beta}_{1,2p-1} = \brac{-1}^{\brac{p+1} / 2} \brac{p-1} \frac{\brac{p-2}!! \brac{3p-4}!!}{4}.
\end{equation}
Comparing this conjectured relation with the explicit values given in \tabref{tabDualBetas}, we find exact agreement.  When $p>5$, this therefore provides highly non-trivial evidence that our conjectured relations, \eqnDref{eqnBeta1,p+n}{eqnBetaRelation}, hold in general (here, $p \leqslant 5$ corresponds to cases in which we have been able to verify our conjectures by explicit computation).

\section{Summary and Conclusions} \label{secConclusion}

The primary fields of a minimal model can be viewed as a complete set of local functionals (which are well-behaved under scale transformations) of the degrees of freedom of the associated critical statistical model (the universality-class representative).  As such, the minimal models provide a complete description of the \emph{local} observables of the critical model. 

The description of boundary effects requires a slight extension of the formalism, in particular, the introduction of boundary-changing operators.  As their name indicates, these operators are considered to be locally inserted at points where the boundary conditions are modified, and are viewed as being responsible for this modification \cite{CarEff86}.  Take for definiteness the Ising model on the upper half plane, with the boundary condition that the spins are all ``$-$'' on the negative real axis and ``$+$'' on the positive real axis.  This can be interpreted as corresponding to inserting a boundary-changing operator, $\func{\psi^{-+}}{0}$ say, at the origin\footnote{We should also remember to insert another boundary-changing operator $\func{\psi^{+-}}{\infty}$ at infinity!} of the theory in which the boundary spins are all negative (say).  Although this insertion operation is local, the action of the inserted operator is inherently non-local.  It changes the sign of \emph{every} spin on the positive real axis.

It should therefore not be surprising that such boundary-changing operators can be related to a certain class of non-local observables.  In particular, crossing probabilities can be expressed in terms of correlators of these operators \cite{CarCri92,SimPer07}.  As noted in the introduction, numerical simulations incorporating these boundary phenomena signal the presence of representations lying outside of the Kac table (which defines the spectrum of minimal models).  This is the general phenomenological framework within which our analysis is anchored.  However, unravelling a direct physical link between the rise of these representations and non-locality is beyond the scope of this article.  Instead, the main question that we tackle is of a technical, but investigative, nature:  How must the irreducible modules that comprise the minimal models (at a chiral level) be modified so as to avoid the decoupling of the representations outside the Kac table from those within?

The starting point of this investigation was the computation of the crossing probability of critical percolation.  From the point of view of \cft{}, the issue was essentially to see how the trivial $\MinMod{2}{3}$ model could be modified to account for a non-trivial $\phi_{1,2}$ four-point function.  In \cite{RidPer07}, we argued that there is only one possible modification which accounts for Cardy's result.  It amounts to replacing the irreducible modules $\IrrMod{1,1}$ and $\IrrMod{1,2}$ (which are identical) by their reducible but indecomposable versions $\IndMod{1,1}$ and $\IndMod{1,2}$.  The theory was then explored by fusing these latter modules repeatedly.  Assuming that such fusing exhausts the spectrum of the theory, this is then the unique consistent (chiral) \cft{} describing critical percolation.  We have shown that it is a logarithmic theory and have denoted it in the present setting by $\LogMinMod{2}{3}$.

Here, we have generalised this ``lifting'' of (chiral) $\MinMod{2}{3}$ to $\LogMinMod{2}{3}$ to all $\MinMod{2}{p}$ models with $p$ odd.  The resulting theories have been again found to be logarithmic, so we have denoted them by $\LogMinMod{2}{p}$.  At the level of representations, the signature of a logarithmic theory lies in the presence of certain indecomposable modules, called staggered modules.  Those that occur in this context are all of rank $2$, and so are fully characterised by the specification of the two \hwms{} from which they are composed, together with the value of the logarithmic coupling that measures the connection between these modules.  These couplings were shown to be independent of the unavoidable ambiguities inherent in fixing the precise form of the logarithmic structure of the staggered modules.  We have phrased such ambiguities in the language of gauge theory.

It should be stressed that these logarithmic couplings, which thus constitute the gauge-invariant data of the staggered modules, are not free parameters, as one might surmise from \cite{RohRed96}.  This is very interesting, as they are frequently quite difficult to compute (the Nahm-Gaberdiel-Kausch fusion algorithm fixes them, but this is not practical in general).  With this in mind, we have observed that they can also be determined by a particular structural feature of the staggered module, specifically by the precise form (indeed, the existence) of the vanishing logarithmic singular vector.  We have computed a number of logarithmic couplings in this way and the equivalence of the results with those of the Nahm-Gaberdiel-Kausch algorithm was extensively checked.  Quite unexpectedly, we were also able to derive explicit closed form expressions (as a function of $p$) for a number of these coupling constants. 
 
However, we must advocate some care with our singular-vector characterisation of the logarithmic coupling.  In a preliminary analysis of the chiral parts of the $\MinMod{1}{p}$ theories\footnote{Despite the notation, these logarithmic theories are not minimal models, and do not cover any minimal model in the way that our $\LogMinMod{2}{p}$ theories do.} \cite{GabInd96}, we have checked (for several cases with $p=2$ and $3$) whether the vanishing logarithmic singular vector of a staggered module determines its logarithmic coupling.  To our surprise, we found that the answer is ``no'':  The vanishing logarithmic singular vector exists for all couplings, and is therefore just a linear function of it.  It is not clear to us at this stage why there should be such a fundamental distinction between these two classes of logarithmic theories, (chiral) $\MinMod{1}{p}$ and $\LogMinMod{2}{p}$.

In the context of \cfts{} with $c=0$, Gurarie and Ludwig \cite{GurCon02,GurCon04} have considered the formulation of an extended theory in which the energy-momentum tensor $T$ has a logarithmic partner $t$ satisfying $\corrfn{\func{T}{z_1} \func{t}{z_2}} = b z_{12}^{-4}$.  The constant $b$ is called there the effective central charge \cite{GurCTh99}.  They have argued that mere global conformal invariance enforces the uniqueness of this charge.  In our terminology, $b$ is the logarithmic coupling of a staggered module whose \hwsm{} is the vacuum module $\IndMod{1,1}$ and whose quotient by this submodule is generated by a dimension $2$ state.  When $c=0$, there are only two such staggered modules:  $\LogMod{1,5}$ and $\LogMod{3,1}$.  As their logarithmic couplings are different, Gurarie and Ludwig's ``unique charge'' argument means that they cannot both be present within a consistent theory.

Given the fusion rules we have uncovered, a more basic statement is that the two indecomposable modules $\IndMod{1,2}$ and $\IndMod{2,1}$ are mutually exclusive.  In the present work, we have proposed a variant of this argument in which the common submodule $\IndMod{1,1}$ of the clashing staggered modules is replaced by $\IndMod{1,2}$.  The conclusion is the same, but this version generalises straight-forwardly, indeed verbatim, to all $\LogMinMod{2}{p}$ (the common submodule is in general $\IndMod{1,p-1}$).  Again, the bottom line is that any fusion ring containing both $\IndMod{1,2}$ and $\IndMod{2,1}$ \emph{cannot}, even if it is consistent as a fusion ring, correspond to any consistent \cft{}, logarithmic or otherwise.

This obstruction not only places severe restrictions on the spectrum of the $\LogMinMod{2}{p}$ model, but it paves the way for a completely different construction of a logarithmic theory rooted in the structure of the $\MinMod{2}{p}$ data.  This ``dual'' theory, which we have denoted by $\DualLogMinMod{2}{p}$, is generated by the indecomposable module $\IndMod{2,1}$.  We have in this way constructed two distinct \emph{logarithmic} \cfts{}, $\LogMinMod{2}{p}$ and $\DualLogMinMod{2}{p}$, with central charge $c = c_{2,p} = 1 - 3 \brac{p-2}^2 / p$, for every odd $p$.  In both cases, we have presented simple and elegant characterisations of their fusion rules.

It is clear that one can generalise this construction still further to consider the two theories generated by modules of the form $\IndMod{1,2}$ or $\IndMod{2,1}$ (respectively) which can be associated to general minimal models $\MinMod{p'}{p}$.  (In principle, this could even be done for irrational $p/p'$.)  However, the physical significance of such theories is not obvious.  In particular, for $p/p' = 4/3$ (corresponding to the Ising model), we would expect to be able to build up a theory which can explain the link, already mentioned in the introduction, between crossing probabilities in the Ising model and the observed exponent $h_{3,3} = \tfrac{1}{6}$.  But neither theory we can construct contains a field of dimension $\tfrac{1}{6}$.  Furthermore, we have checked that the obstruction to including both $\IndMod{1,2}$ or $\IndMod{2,1}$ in a consistent theory is still present in this case, so it is not possible to generate a field of dimension $\tfrac{1}{6}$ in this way either.  We hope to report on the resolution of this puzzle in the future.

For our final comment, we return to critical percolation.  As has already been stressed, the expression for the crossing probability has been rigorously demonstrated by mathematicians \cite{LawValI01,SmiCri01}.  This proof is based on \emph{Schramm-L\"{o}wner Evolution} (SLE) techniques (see for example the review \cite{CarSLE05}), and verifies that percolation corresponds to $\kappa = 6$, where $\kappa$ is the value of the SLE diffusion constant.  On the other hand, we have provided a consistent field-theoretic framework for Cardy's original determination of the crossing probability, the model $\LogMinMod{2}{3}$.  It is thus natural to propose the equivalence:
\begin{equation}
\text{SLE}_6 \sim \LogMinMod{2}{3}.
\end{equation}
We therefore hold that this SLE model is in fact equivalent to a \emph{logarithmic} \cft{}.  Similarly, by duality, we have already conjectured that the other $c=0$ statistical model, self-avoiding walks, is also described by a logarithmic theory, our $\DualLogMinMod{2}{3}$.  The natural conjecture in this case is then that
\begin{equation}
\text{SLE}_{8/3} \sim \DualLogMinMod{2}{3}.
\end{equation}
These conjectural equivalences and their generalisations will be considered in more detail elsewhere.

\section*{Acknowledgements}

We thank Yvan Saint-Aubin for valuable correspondence whilst preparing this article.  DR also thanks Stefan Fredenhagen, Volker Schomerus, J\"{o}rg Teschner, and Stefan Theisen for interesting discussions.

\appendix

\section{Adjoints of Fields} \label{appFieldAdjoints}

In this appendix, we justify the sign appearing in \eqnref{eqnCF23:12,32}.  This is straight-forward, but involves a subtlety which we feel deserves to be addressed explicitly.  This subtlety does not appear in ordinary \cft{}, and arises from the presence of non-vanishing singular vectors in the theory.

The general solution of the \ode{} (\ref{eqnODEchilam}) is easily checked to be
\begin{equation}
\corrfn{\func{\chi_{1,2}}{z_1} \func{\lambda_{3,2}}{z_2}} = \frac{C}{z_{12}^2},
\end{equation}
where $C$ is an undetermined constant.  As opposed to the constant appearing in \eqnref{eqnCorrFn23:14,32}, this constant is clearly gauge-invariant.  We therefore expect that it can be related to the logarithmic coupling.  Indeed, since $\chi_{1,2}$ is primary with dimension $1$, we might expect to be able to compute $C$ as
\begin{equation} \label{eqnWrong}
C = \lim_{\substack{z_1 \rightarrow \infty \\ z_2 \rightarrow 0}} \ \frac{C z_1^2}{z_{12}^2} = \lim_{\substack{z_1 \rightarrow \infty \\ z_2 \rightarrow 0}} \ z_1^2 \corrfn{\func{\chi_{1,2}}{z_1} \func{\lambda_{3,2}}{z_2}} = \braket{\chi_{1,2}}{\lambda_{3,2}} = \beta_{3,2}.
\end{equation}

However, we might also recall that $\chi_{1,2} = \partial \phi_{1,2}$, and hence try to compute $\corrfn{\func{\chi_{1,2}}{z_1} \func{\lambda_{3,2}}{z_2}}$ by differentiating $\corrfn{\func{\phi_{1,2}}{z_1} \func{\lambda_{3,2}}{z_2}}$ with respect to $z_1$.  The latter correlator solves the following \ode{} induced by the global conformal invariance generated by $L_{-1}$ and $L_0$:
\begin{equation}
\brac{z_{12} \diff{}{z_{12}} + 1} \corrfn{\func{\phi_{1,2}}{z_1} \func{\lambda_{3,2}}{z_2}} = -\corrfn{\func{\phi_{1,2}}{z_1} \func{\chi_{1,2}}{z_2}} = 0,
\end{equation}
hence has the form $C' / z_{12}$.  This $C'$ is again gauge-invariant, and is determined by the \pde{}
\begin{equation}
\brac{z_1^2 \pardiff{}{z_1} + z_2^2 \pardiff{}{z_2} + 2 z_2} \corrfn{\func{\phi_{1,2}}{z_1} \func{\lambda_{3,2}}{z_2}} = -\beta_{3,2} \corrfn{\func{\phi_{1,2}}{z_1} \func{\phi_{1,2}}{z_2}} = -\beta_{3,2},
\end{equation}
corresponding to $L_1$-global conformal invariance.  Substituting our general solution into this gives $C' = \beta_{3,2}$.  But if we now differentiate $\corrfn{\func{\phi_{1,2}}{z_1} \func{\lambda_{3,2}}{z_2}}$, we obtain
\begin{equation} \label{eqnCorrect}
\corrfn{\func{\chi_{1,2}}{z_1} \func{\lambda_{3,2}}{z_2}} = \frac{-\beta_{3,2}}{z_{12}^2},
\end{equation}
that is, $C = -\beta_{3,2}$.  We therefore have a contradiction to resolve.

The resolution is that our first computation (\ref{eqnWrong}) is incorrect.  More specifically, we cannot conclude that
\begin{equation}
\bra{\chi_{1,2}} = \lim_{z_1 \rightarrow \infty} z_1^2 \bra{0} \func{\chi_{1,2}}{z_1},
\end{equation}
just because $\chi_{1,2}$ is primary.  Such a definition, which really defines the \emph{adjoint} of this field, should be restricted to (primary) generators of the module.  Once the adjoint is defined for generating fields, the adjoints of the descendants are completely fixed, regardless of whether they happen to be primary or not.

To illustrate this, let us note that the adjoint of a generating primary field $\phi$ of dimension $h$ merely reflects the adjoint operation on its modes:
\begin{equation}
\func{\phi}{z} = \sum_n \phi_n z^{-n-h} \qquad \Rightarrow \qquad \func{\phi}{z}^{\dag} = \sum_n \phi_n^{\dag} z^{-n-h}.
\end{equation}
If we define the adjoint of $\func{\phi}{z}$ in the usual way, then
\begin{equation}
\bra{\phi} = \lim_{z \rightarrow \infty} \ z^{2h} \bra{0} \func{\phi}{z} = \lim_{z \rightarrow 0} z^{-2h} \ \bra{0} \func{\phi}{z^{-1}} = \lim_{z \rightarrow 0} \ \bra{0} \sum_n \phi_n z^{n-h} = \lim_{z \rightarrow 0} \ \bra{0} \sum_n \phi_{-n} z^{-n-h},
\end{equation}
so we see that this definition is really equivalent to $\phi_n^{\dag} = \phi_{-n}$.  The modes of $\func{\partial \phi}{z}$ therefore satisfy
\begin{equation}
\brac{\partial \phi}_n = \brac{-n-h} \phi_n \qquad \Rightarrow \qquad \brac{\partial \phi}_n^{\dag} = \brac{-n-h} \phi_{-n},
\end{equation}
hence
\begin{align}
\bra{\partial \phi} &= \lim_{z \rightarrow 0} \ \bra{0} \func{\partial \phi}{z}^{\dag} = \lim_{z \rightarrow 0} \ \bra{0} \sum_n \brac{-n-h} \phi_{-n} z^{-n-h-1} = \lim_{z \rightarrow 0} \ \bra{0} \sum_n \brac{n-h} \phi_n z^{n-h-1} \notag \\
&= \lim_{z \rightarrow 0} \ \bra{0} \sqbrac{z^{-2h-2} \sum_n \brac{n+h} \phi_n z^{n+h+1} - 2h z^{-2h-1} \sum_n \phi_n z^{n+h}} \notag \\
&= -\lim_{z \rightarrow 0} \ \bra{0} \sqbrac{z^{-2h-2} \func{\partial \phi}{z^{-1}} + 2h z^{-2h-1} \func{\phi}{z^{-1}}} \notag \\
&= -\lim_{z \rightarrow \infty} \ \sqbrac{z^{2h+2} \bra{0} \func{\partial \phi}{z} + 2h z^{2h+1} \bra{0} \func{\phi}{z}}.
\end{align}

Returning to \eqnref{eqnWrong}, we can now correctly write
\begin{equation}
C = \lim_{\substack{z_1 \rightarrow \infty \\ z_2 \rightarrow 0}} \ \frac{C z_1^2}{z_{12}^2} = \lim_{\substack{z_1 \rightarrow \infty \\ z_2 \rightarrow 0}} \ z_1^2 \corrfn{\func{\partial \phi_{1,2}}{z_1} \func{\lambda_{3,2}}{z_2}} = -\braket{\partial \phi_{1,2}}{\lambda_{3,2}} = -\beta_{3,2},
\end{equation}
in full agreement with \eqnref{eqnCorrect}.  The sign in \eqnref{eqnCF23:12,32} is therefore as given.

\section{Mutual Locality} \label{appMutualBosons}

Two fields $\phi$ and $\psi$ are said to be \emph{mutually local} if
\begin{equation}
\func{\phi}{z} \func{\psi}{w} = \mu \func{\psi}{w} \func{\phi}{z}
\end{equation}
for some $\mu \neq 0$.  This (or an appropriate generalisation) is practically axiomatic in \cft{}.  Such fields are mutually \emph{bosonic} if $\mu = 1$ and mutually \emph{fermionic} if $\mu = -1$.  It is generally assumed, but rarely (if ever) proved, that Virasoro primaries are mutually bosonic.  We will show that $T$ and any Virasoro primary $\phi$ are mutually bosonic, not because this is particularly interesting itself\footnote{It is much more interesting when the symmetry algebra is affine.  For example, (chiral) primaries of the $\func{\group{SU}}{2}$ Wess-Zumino-Witten model are generally mutually bosonic with the field corresponding to the maximal torus, but mutually fermionic with the other affine fields \cite{RidSU206}.}, but because a simple corollary of this is that the statement remains true when $\phi$ is replaced by a logarithmic partner field $\lambda$.

Suppose therefore that $\phi$ is primary of dimension $h$ and that
\begin{equation}
\func{T}{z} \func{\phi}{w} = \mu \func{\phi}{w} \func{T}{z}.
\end{equation}
Defining the generalised commutator
\begin{equation}
\dcomm{L_m}{\phi_n} = L_m \phi_n - \mu \phi_n L_m, \qquad \dcomm{\phi_n}{L_m} = \phi_n L_m - \mu^{-1} L_m \phi_n = - \mu^{-1} \dcomm{L_m}{\phi_n},
\end{equation}
the standard \ope{} gives
\begin{equation} \label{eqnGenCommPrim}
\dcomm{L_m}{\phi_n} = \bigl( m \brac{h-1} - n \bigr) \phi_{m+n}.
\end{equation}
We have to satisfy the Jacobi identity for this generalised commutator, in particular,
\begin{equation}
\dcomm{L_m}{\dcomm{L_n}{\phi_p}} - \dcomm{L_n}{\dcomm{L_m}{\phi_p}} = \mu \dcomm{\comm{L_m}{L_n}}{\phi_p}.
\end{equation}
Substituting \eqnref{eqnGenCommPrim} into this identity and simplifying gives
\begin{equation}
\brac{\mu - 1} \brac{m-n} \bigl( \brac{m+n} \brac{h-1} -p \bigr) \phi_{m+n+p} = 0,
\end{equation}
whence $\mu = 1$.

Let us now replace $\phi$ by an arbitrary (generating) logarithmic partner field $\lambda$ of dimension $h'$.  The relevant \ope{} takes the form
\begin{equation}
\func{T}{z} \func{\lambda}{w} = \ldots + \frac{h' \func{\lambda}{w} + \func{\chi}{w}}{\brac{z-w}^2} + \frac{\func{\partial \lambda}{w}}{z-w} + \ldots,
\end{equation}
where the omitted singular terms only involve the corresponding primary $\phi$ and its descendants.  Indeed, the (singular) terms involving $\lambda$ are precisely what one would expect if $\lambda$ were primary.  When we repeat the above analysis to determine the mutual locality coefficient $\mu$ for $T$ and $\lambda$, the Jacobi identity must therefore reduce to the form
\begin{equation} \label{eqnmu=1}
\brac{\mu - 1} \brac{m-n} \bigl( \brac{m+n} \brac{h'-1} -p \bigr) \lambda_{m+n+p} + \ldots = 0,
\end{equation}
where the omitted terms correspond to the mode $\phi_{m+n+p}$ and modes of the descendants of $\phi$.  It follows again that $\mu$ must be $1$ --- logarithmic partner fields are also mutually bosonic with $T$.  One can of course check, at least in specific cases, that the omitted terms in \eqnref{eqnmu=1} also vanish when $\mu = 1$.

As a quick application of this result, let us use this to prove that $\phi_{1,2}$ and $\lambda_{1,4}$ are mutually bosonic in $\LogMinMod{2}{3}$ (this result is needed to justify \eqnref{eqnCF23:14,12}).  Let us suppose then that
\begin{equation} \label{eqnMutLoc}
\func{\phi_{1,2}}{z} \func{\lambda_{1,4}}{w} = \mu \func{\lambda_{1,4}}{w} \func{\phi_{1,2}}{z}.
\end{equation}
We expand both sides of
\begin{equation}
\corrfn{\func{T}{x} \func{\phi_{1,2}}{z} \func{\lambda_{1,4}}{w}} = \mu \corrfn{\func{T}{x} \func{\lambda_{1,4}}{w} \func{\phi_{1,2}}{z}}
\end{equation}
using the \opes{} of $T$ with $\phi_{1,2}$ and $\lambda_{1,4}$ (the latter is given in \eqnref{eqnOPETlam}), and the fact that $T$ is mutually bosonic with these fields.  Comparing the terms which result, most give no constraint on $\mu$.  For example, the coefficients of $\brac{x-z}^{-1}$ give
\begin{equation}
\partial_z \corrfn{\func{\phi_{1,2}}{z} \func{\lambda_{1,4}}{w}} = \mu \partial_z \corrfn{\func{\lambda_{1,4}}{w} \func{\phi_{1,2}}{z}},
\end{equation}
in agreement with \eqnref{eqnMutLoc} for arbitrary $\mu$.  However, the terms which arise from the logarithmic nature of $\lambda_{1,4}$ are more interesting.  In this example the $\brac{x-w}^{-3}$ coefficients give
\begin{equation}
\beta_{1,4} \corrfn{\func{\phi_{1,2}}{z} \func{\phi_{1,2}}{w}} = \mu \beta_{1,4} \corrfn{\func{\phi_{1,2}}{w} \func{\phi_{1,2}}{z}} \qquad \Rightarrow \qquad \mu = 1,
\end{equation}
since $\corrfn{\func{\phi_{1,2}}{z} \func{\phi_{1,2}}{w}} = 1$ (and $\beta_{1,4} \neq 0$).  This proves that $\phi_{1,2}$ and $\lambda_{1,4}$ are mutually bosonic as required.  We remark that this simple proof would not be valid if $\lambda_{1,4}$ were primary.  The logarithmic nature of this field actually makes the analysis easier than in the standard case!

\section{Inner Products for Staggered Modules} \label{appInnProd}

As in \cite{RidPer07} (albeit implicitly there), we fix the inner product on the staggered modules by setting the norm of the \hws{} $\ket{\phi}$ of the maximal \hwsm{} to unity.  This has at least one advantage from a mathematical perspective:  This restricts to the usual inner product on this maximal submodule, so it can be treated as a module on its own terms.  There is, however, at least one disadvantage:  As shown in \cite{RidPer07} (see footnote~8 and the remark after Equation~(3.15)), the norms of the logarithmic generating states $\ket{\lambda}$ in the staggered modules must \emph{diverge}.  Nevertheless, the inner product of the non-vanishing singular vector $\ket{\chi}$ and its logarithmic partner $\ket{\lambda}$ is \emph{finite}, and in fact defines the logarithmic coupling $\beta$ of the module (\eqnref{eqnDefLogCoupling}).  At the level of fields, where the r\^{o}le of the inner product is taken over by the $2$-point functions, it is easy to check that our choice of inner product gives a well-defined \emph{non-degenerate} matrix of $2$-point ``constants'' (actually functions).

Of course, this is not the only inner product one could choose.  At first glance, one might think that defining the norm of $\ket{\lambda}$ to be $1$ will be more useful.  After all, a rank $2$ staggered module is generated by this state.  This has the immediate advantage that every state in the module now has finite norm.  However, that of the \hws{} $\ket{\phi}$ is now necessarily zero.  Thus, $\ket{\phi}$ is null, $\ket{\chi}$ is null, and (as we shall see below) the overlap $\braket{\chi}{\lambda}$ also vanishes.  It follows that this inner product leads to degeneracy in \emph{all} $2$-point constant matrices, a mathematically unpleasant situation.

From the point of view of the representation theory, there is no canonical choice.  Mathematically, one can do as one wishes, provided one does not encounter a contradiction\footnote{In the somewhat special case of a Jordan \hwm{}, in which the \hws{} $\ket{\phi}$ itself has a Jordan partner $\ket{\lambda}$, one can derive such a contradiction.  For these modules, we have to take $\ket{\phi}$ to be null.  Viewed as an extreme case of a staggered module, this is natural in that the null state $\ket{\chi}$ would coincide with $\ket{\phi}$.  But note that Jordan \hwms{} do not appear in the models under investigation here.}.  But in \cft{}, such a choice should (ideally) be grounded in physical considerations.  This is indeed the case here:  The following two propositions demonstrate that, at least for the $\LogMinMod{2}{3}$ model, our choice is forced by the non-triviality of Cardy's $4$-point function.

\begin{proposition}
If the $\LogMinMod{2}{3}$ \hws{} $\ket{\phi_{1,2}}$ is null, then Cardy's formula for the horizontal crossing probability of critical percolation (as the $4$-point function of $\func{\phi_{1,2}}{z}$) vanishes identically.
\end{proposition}
\begin{proof}
Since $\phi_{1,2}$ has dimension $0$, $\ket{\phi_{1,2}}$ null implies that
\begin{equation}
\corrfn{\func{\phi_{1,2}}{z} \func{\phi_{1,2}}{w}} = \braket{\phi_{1,2}}{\phi_{1,2}} = 0.
\end{equation}
Moreover, since the fusion algorithm of Nahm and Gaberdiel-Kausch does not care whether states are null or not,
\begin{align}
L_1 \ket{\lambda_{1,4}} &= \beta_{1,4} \ket{\phi_{1,2}} & \text{(algebraic fusion algorithm)} \\
\Rightarrow \qquad \braket{\chi_{1,2}}{\lambda_{1,4}} &= \beta_{1,4} \braket{\phi_{1,2}}{\phi_{1,2}} = 0 & \\
\Rightarrow \qquad \corrfn{\func{\chi_{1,2}}{z} \func{\lambda_{1,4}}{w}} &= 0 & \text{(global conformal invariance)} \\
\Rightarrow \qquad \corrfn{\func{\phi_{1,2}}{z} \func{\lambda_{1,4}}{w}} &= 0 & \text{($\func{\chi_{1,2}}{z} = \func{\partial \phi_{1,2}}{z}$).}
\end{align}
Indeed, this last correlation function could be an arbitrary function of $w$ alone (the $z$ derivative being zero), but we already know that it is actually a function of $z-w$, hence it must vanish.

Now consider Cardy's $4$-point function $\corrfn{\func{\phi_{1,2}}{z_1} \func{\phi_{1,2}}{z_2} \func{\phi_{1,2}}{z_3} \func{\phi_{1,2}}{z_4}}$.  Knowing the fusion rule
\begin{equation}
\IndMod{1,2} \fuse \IndMod{1,2} \fuse \IndMod{1,2} = \IndMod{1,2} \oplus \LogMod{1,4},
\end{equation}
we can reduce this $4$-point function to an infinite linear combination of ``descendant correlators'' of
\begin{equation}
\corrfn{\func{\phi_{1,2}}{z_1} \func{\phi_{1,2}}{z_4}} \qquad \text{and} \qquad \corrfn{\func{\phi_{1,2}}{z_1} \func{\lambda_{1,4}}{z_4}}.
\end{equation}
But, we have seen that both these correlators vanish when $\ket{\phi_{1,2}}$ is null, hence so do all their descendant correlators, and the vanishing of Cardy's $4$-point function follows.
\end{proof}

\begin{proposition}
If the vacuum $\ket{0}$ is null, so is $\ket{\phi_{1,2}}$.
\end{proposition}
\begin{proof}
Using $\IndMod{1,2} \fuse \IndMod{1,2} = \IndMod{1,1} \oplus \IndMod{1,3}$, we can reduce $\corrfn{\func{\phi_{1,2}}{z} \func{\phi_{1,2}}{w}}$ to an infinite linear combination of descendant correlators of
\begin{equation}
\corrfn{\func{\phi_{1,1}}{w}} = \braket{0}{0} \qquad \text{and} \qquad \corrfn{\func{\lambda_{1,3}}{w}} = 0.
\end{equation}
If $\ket{0}$ is null, it follows that $\braket{\phi_{1,2}}{\phi_{1,2}} = \corrfn{\func{\phi_{1,2}}{z} \func{\phi_{1,2}}{w}} = 0$.
\end{proof}

\noindent The contrapositives of these propositions together prove that a null vacuum state implies that Cardy's $4$-point function vanishes identically (in $\LogMinMod{2}{3}$).  This justifies, physically, our choice of an inner product in which the vacuum is \emph{not} null, and suggests that the same choice will be the physically relevant one in the other $\LogMinMod{2}{p}$ models.  We mention that in many other logarithmic theories, the $\MinMod{1}{p}$ theories in particular, one is forced to take an inner product for which the vacuum is null, as the vacuum has a non-trivial Jordan partner.  The above propositions therefore also prove that the possession of a null vacuum state is not a necessary condition of logarithmic \cft{}.

\end{document}